\documentclass{elsart} %   % Elsevier article

\usepackage{graphicx}    % for including eps-figure files                      
\usepackage{amssymb,amsmath,amsbsy}
\usepackage[T1]{fontenc}

\newcommand{\greeksym}[1]{{\usefont{U}{psy}{m}{n}#1}}

\def\slantfrac#1#2{\kern.1em^{#1}\kern-.3em/\kern-.1em_{#2}}
\def\b#1{\mathbf{#1}}

\def\m#1{\mathrm{#1}}
\def\slantfrac#1#2{\kern.1em^{#1}\kern-.3em/\kern-.1em_{#2}}
\providecommand\lesssim{\mathrel{\mathpalette\vereq<}}
\providecommand\gtrsim{\mathrel{\mathpalette\vereq>}}

\begin{document}
\begin{frontmatter}
\title{Critical velocities in superfluids and the nucleation of vortices}
\author{\'Eric Varoquaux}
\address{CNRS--Universit\'{e} Paris-Sud,
     Laboratoire de Physique des Solides,\\
     B\^atiment 510, F-91405 Orsay Cedex, France\\
     and \\
     Commissariat \`a l'\'Energie Atomique, 
     Service de Physique de l'\'Etat Condens\'e, \\
     B\^atiment 772, Centre de Saclay, F-91191 Gif-sur-Yvette Cedex, France
}
%n
\begin{abstract}
  The problem of critical velocities in superfluids, that is the comprehension
  of superfluidity breakdown by flow, has been long standing. One difficulty
  stems from the existence of several breakdown mechanisms. A major advance
  has come from the observation of single $2\pi$ phase slips, which arise from
  the nucleation of quantised vortices, that is, their creation {\it ex
    nihilo}. The statistical properties of the nucleation process in both the
  thermal regime and the quantum regime are identified and analysed: vortex
  nucleation provides a well-documented case of macroscopic quantum tunnelling
  (MQT). In particular, a close scrutiny of the experimental data obtained on
  ultra-pure $^4$He reveals the influence of damping on tunnelling, a rare
  occurrence where the effect of the environment on MQT can be studied.
  
  La vitesse critique dans les superfluides, c'est à dire la destruction de la
  superfluidité par l'écoulement du fluide, pose un problème qui perdure. Une
  des difficultés réside dans l'existence de plusieurs mécanismes pour cette
  destruction. L'observation de sauts de phase individuels de $2\pi$, qui
  proviennent de la nucléation de tourbillons quantifiés, a constitué une
  avancée importante. L'identification et l'analyse des propriétés
  stochastiques du processus de nucléation, tant dans le régime classique que
  quantique, ont conduit à l'étude très circonstanciée d'un cas spécifique
  d'effet tunnel macroscopique. En particulier, l'examen fouillé des données
  expérimentales obtenues avec l'hélium ultra-pur a révélé l'influence de
  la dissipation sur l'effet tunnel, donnant par là un exemple rare
  d'interaction d'un processus tunnel avec son environnement macroscopique.

\end{abstract}
\end{frontmatter}

The critical velocity in a superfluid is the threshold above which the flow of
the superfluid component becomes dissipative, that is, the property of
superfluidity is lost. This rather broad definition encompasses a number of
different physical situations. The following overview starts with a brief
description of the different brands of velocities that comply with this
definition. It then joins the main trend of this Dossier by focusing on
that which involves a nucleation phenomenon, namely, the nucleation of
superfluid vortices.

Neither the problem of critical velocities in superfluids nor that of the
nucleation of vortices are new. The former is as old as the discovery of
superfluidity (see the monograph by Wilks \cite{Wilks:67}). The latter, first
discussed by Vinen in the early sixties \cite{Vinen:63}, has met a more
tortuous fate. It was first thought, still is in some quarters, to be
impossible \cite{Vinen:63} on the grounds that such an extended hydrodynamical
object as a vortex with a finite (quantised) circulation, involving the motion
of a large number of helium atoms, would have a vanishingly small probability
of occurring spontaneously. For classical ideal fluids, this is the essence of
the Kelvin-Helmholtz theorem, which states that vorticity is conserved for
isentropic motion of inviscid fluids. More recent experiments,
probing superflow on a finer scale of length
\cite{Langer:70,Muirhead:84,Varoquaux:87}, have shown otherwise.

I give below a short account of these problems, which is hardly more
than a guided tour of the four references
\cite{Varoquaux:91,Avenel:93,Varoquaux:01,Varoquaux:03} written by the author
and his colleagues over the course of many years. More extended discussions
can be found in these articles as well as more complete bibliographies. Also,
a more comprehensive review is in preparation.

% %%%%%%%%%%%%%%%%%%%%%%%%%%%%%%%%%%%%%%%%%%%%%% 
% 
\section{Critical velocities in superfluids} 
% 
%%%%%%%%%%%%%%%%%%%%%%%%%%%%%%%%%%%%%%%%%%%%%%% 

\subsection{The Landau criterion}

% ---------------------------------------------------------------------------
\begin{figure}[t]
  \begin{center}
    \includegraphics[height=57mm]{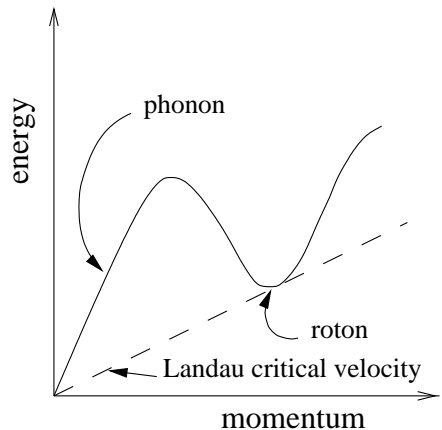}
    \caption{\label{DispersionCurve} Dispersion curve of the elementary
      excitations in superfluid $^4$He }
  \end{center}
\end{figure}
% ---------------------------------------------------------------------------

Landau \cite{Wilks:67} explained the superfluidity of helium-4 by the
sharpness of the dispersion curve for elementary excitations, phonons and
rotons, shown in Fig.\,\ref{DispersionCurve}, which is a property associated
with the existence of a Bose-Einstein condensate. Elementary excitation energy
levels $\epsilon(p)$ being well-defined, that is, having a negligible spread
in energy, very low-lying states, energy-wise and momentum-wise, are extremely
scarce.  An impurity, or a solid obstacle, can only exchange an
energy $\epsilon(p)$ at momentum $p$ that exactly matches the energy of an
elementary excitation of the fluid. Unless this condition can be precisely
met, there is no dissipative interaction between the fluid and its
surroundings: the flow is viscousless at small flow velocities.

If the superfluid moves at velocity $\b v_\m s$, the energy of elementary
excitations in the frame of reference at rest becomes $\epsilon+\b v_\m s\cdot
\b p$ \cite{Wilks:67,Baym:69}. The same holds for a moving obstacle, by
Galilean invariance. If this energy turns negative, elementary excitations
proliferate and superfluidity is lost. The condition on the superfluid
velocity for this to happen reads:
\begin{equation}        \label{LandauVelocity}
  v_\m s \geqslant  v_\m L = \frac{\epsilon(p)}{p}\Big|_\m{min} \simeq 
    \frac{\epsilon(p)}{p}\Big|_\m{roton} \; .
\end{equation}

The minimum value of $\epsilon/p$ for helium is very close to the roton
minimum, as shown in Fig.\,\ref{DispersionCurve}. In $^4$He at low pressure,
$v_\m L \simeq 60$ m/s. The Landau critical velocity $v_\m L$ is smaller than
the sound velocity $c=220$ m/s but larger than most critical velocities
measured in various experiments. For the much less dense Bose-Einstein
Condensed gases, which do not exhibit roton-like features, the minimum is the
sound velocity, $c=\epsilon(p)/p|_{p=0}$.

%%%%%%%%%%%%%%%%%%%%%%%%%%%%%%%%%%%%%%%%%%%
\subsection{Feynman's approach}

%%%%%%%%%%%%%%%%%%%%%%%%%%%%%%%%%%%%%%%%%%%

Feynman \cite{Wilks:67}, following Onsager, realised that, not only would
vorticity be quantised in $^4$He in units of the quantum of circulation
$\kappa_4=2\pi \hbar/m_4\simeq 10^{-3}$ cm$^2$/s, $m_4$ being the mass of the
helium-4 atom (which is also a property associated with the existence of a
Bose-Einstein condensate) but that these vortices would be responsible for the
onset of dissipation and for a critical velocity in the superfluid. The basic
reason for this, as spelled out clearly by Anderson \cite{Anderson:66a}, is
that vortices can exchange energy with the potential superflow and carry this
energy away, thus causing an energy loss to the superflow.

In order to evaluate a characteristic velocity associated with this process,
let us consider a vortex ring of radius $R$. Its energy $E_\m R$ and impulse
$P_\m R$ are expressed by
\cite{Lamb:32,Sonin:87,Sonin:95} 
\begin{equation}        \label{RingEnergy}
 E_\m R = \frac{1}{2}\rho_\m s \kappa_4^2 R\,\left(\ln \frac{R}{a_0} -
    \frac{7}{4}\right) + {\mathcal O}\left(\frac{a_0}{R}\right) \; .
\end{equation}
\begin{equation}        \label{RingImpulse}
  P_\m R =  \pi \rho_\m s \kappa_4 R^2 \; ,
\end{equation}
where $a_0$ is the vortex core radius, taken here as the superfluid coherence
length.

Let us treat such a vortex as an elementary excitation of the superfluid,
which it rightfully is, and apply Landau's criterion. The limiting velocity is
reached for a radius $R$ such that $E_\m R/P_\m R$ is at a minimum, which
occurs when $R$ is as large as feasible, that is of the order of the channel
size $d$. This minimum value sets the velocity at which vortices can start to
appear and defines the Feynman critical velocity:
\begin{equation}        \label{FeynmanVelocity}
  v_\m F \simeq \frac{\kappa_4}{2\pi d} \ln\left(\frac{d}{a_0}\right) \; .
\end{equation}
As discussed below, $v_\m F$ is much closer to experimental values than the
Landau critical velocity for rotons. Although this agreement is heartening, it
also raises fresh questions: how do these vortices come about? 

% %%%%%%%%%%%%%%%%%%%%%%%%%%%%%%%%%%%%%%%%%%%
% \subsection{Hydrodynamic instabilities}

% %%%%%%%%%%%%%%%%%%%%%%%%%%%%%%%%%%%%%%%%%%%

%  Kelvin-Helmholtz -> non-isentropic process, or
% breakdown of classical hydrodynamic (quantum effects)

% Ricca-Frisch-Pomeau, using the Gross-Pitaevski equation. Vortices are emitted.

% But critical velocity value is wrong (right for BEC). Because no rotons in the G-P
% model, phonon velocity. Purport the Landau criterion, but vortices instead of
% phonons. Lambert for improvements.

% If G-P model is improved to include rotons (Huepe, Caroiline Nore), rotons are
% emitted, not vortices! In fact, this correponds to experiments with ions
% (Bowley, McClintock). Figure to illustrate.

% However, in capillaries and apertures, 1) critical velocities much lower than
% $v_\m L$, 2) no clear pattern emerging at first sight, at least until a
% process involving vortices could be clearly seen in aperture flows. 

\subsection{The phase slips}

% ---------------------------------------------------------------------------
\begin{figure}[t]
  \begin{center}
    \includegraphics[height=45 mm,angle=0]{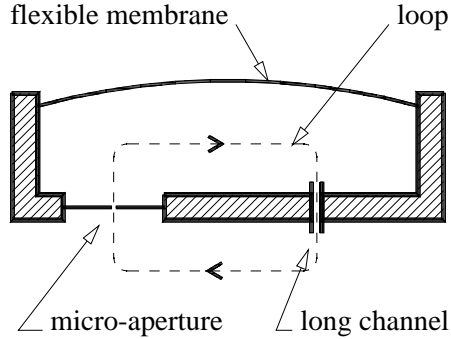}
    \caption{\label{Cell}Schematic drawing of the flexible-diaphragm Helmholtz
      resonator. }
  \end{center}
\end{figure}
% ---------------------------------------------------------------------------

The phase slippage experiments that were carried out starting from the
mid-eighties \cite{Varoquaux:87,Avenel:85} confirmed Feynman and Anderson's
views on dissipation in superflows \cite{Anderson:66a} and brought a large
measure of clarification in the critical velocity problem \cite{Varoquaux:91}
and in the formation of vortices in superfluid $^4$He \cite{Avenel:93}. These
experimental results and their interpretation have since been largely confirmed
\cite{Shifflett:92,Zimmermann:96,Packard:98}

Phase slips can be studied with the help of a miniature hydro-mechanical
device, which is basically a flexible-diaphragm Helmholtz resonator as
represented schematically in Fig.\,\ref{Cell}. This resonator is immersed in the
superfluid bath. The flexible diaphragm is constituted by a Kapton membrane
coated with aluminium. In the version shown in Fig.\,\ref{Cell}, there are two
openings connecting the resonator chamber to the superfluid bath. One is a
micro-aperture in which the critical velocity phenomenon takes place.  The
critical event consists in a sudden jump in the resonance amplitude which
corresponds to an abrupt change in the flow velocity through the
micro-aperture and a loss of resonator energy. These dissipation events are
interpreted as resulting from single vortex emission, to which is associated a
slip by $2\pi$ of the quantum phase difference across the micro-aperture,
$\delta \varphi$, caused by the motion of the vortex across the flow stream.

The other opening is a relatively open duct and provides a parallel path to
the superfluid along which the quantum phase remains well determined even when
the phase slips in the micro-aperture.  A quantum of circulation builds up for
each $2\pi$ slip along the superfluid closed loop threading the two openings.
The operation of these resonators is described in detail in the
literature (see, for instance
\cite{Varoquaux:87,Avenel:87,Beecken:87,Varoquaux:94,Avenel:95}).

% ---------------------------------------------------------------------------
\begin{figure}[t]
  \begin{center}
    \hskip -6 mm
    \includegraphics[height=14.5 cm ,angle=90]{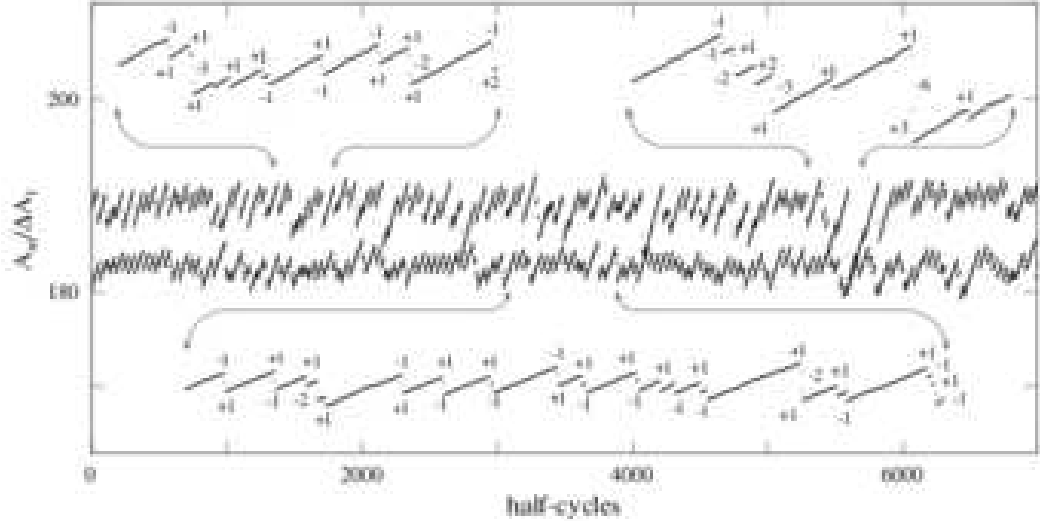}
    \caption{\label{TimeChart}Peak amplitude charts at 100.7 mK (top) and
      201.5 mK (bottom) at saturated vapour pressure in ultra-pure $^4$He with
      a $^3$He impurity concentration less than 10$^{-9}$. Peak absolute
      amplitudes during each half-cycle of the resonator motion at 12.5 Hz are
      plotted as a function of half-cycle index; time runs from left to right.
      The peak membrane amplitude $A_\m m$ is normalised to the amplitude jump of
      a single slip $\Delta A_1$. The traces at the very top and bottom of the
      graph, expanded from the main traces as indicated by curly arrows, show
      the slip sizes in signed winding numbers of the quantum phase (according
      to flow direction, in and out of the resonator chamber). }
  \end{center}
\end{figure}
% ---------------------------------------------------------------------------

The resonator is driven on resonance by an electrostatic {\it ac}-drive
applied to the aluminium-coated flexible membrane at a constant level. In the
absence of dissipation, the resonance motion increases linearly in amplitude
under the action of the drive. The displacement of the membrane, which is
proportional to the total flow in and out of the resonator, is monitored by an
electrodynamic displacement sensor with a SQUID amplifier \cite{SQUID}, the
output of which is converted into a digital signal by a fast
analogue-to-digital converter.  The peak amplitudes of the membrane motion
$A_\m m$ for each successive positive-going and negative-going half-cycle of the
resonance are measured and stored digitally in real time.  The raw data
consist of peak amplitude charts as represented in Fig.\,{\ref{TimeChart}}.
Phase slips are seen as sudden drops of the peak amplitude from one
half-cycle to the next.

The pattern of these peak amplitude data is processed numerically to identify
all the slips and their multiplicity, and to obtain the state of quantised
circulation trapped in the resonator loop, labelled $j$. The actual flow in
the micro-aperture is the sum of the flow driven by the membrane and of the
persistent flow threading the micro-aperture and the parallel channel, which
depends on the quantum state of the loop $j$.  The amplitude drop $\Delta A_1$
caused by a single phase slip in a given half-cycle of the resonance
corresponds to a change of $\delta \varphi$ by exactly $2\pi$.  Normalising
the membrane displacement $A_\m m$ by $\Delta A_1$ as done in
Fig.\,\ref{TimeChart} provides a self-calibration of the data that is
independent of less well known quantities such as the membrane stiffness, and
the calibration factor of the displacement sensor.

%  The peak aperture velocity $v_p$ is deduced from the
% peak membrane amplitude and from the trapped circulation by
% \cite{Avenel:95,Avenel:98}
% %
% \begin{equation}         \label{WindingNumber}
% v_p = \frac{A_m}{\Delta A_1(1+R)} + (j+\frac{\kappa_b}{\kappa_4})\,
%       \frac{R}{1+R}  \;.
% \end{equation}
% %
% In (\ref{WindingNumber}) and
 
In the following, aperture velocities are expressed by the number of turns
$\delta \varphi /2\pi$ by which the quantum-mechanical phase winds across the
aperture \cite{winding}. The actual flow velocity averaged over the
cross-section of the micro-aperture is proportional to $\delta \varphi$, the
multiplying factor being $\hbar /m_4 l_\m h$.  The hydraulic length $l_\m h$
of the micro-aperture is of the order of 1
{\small\raisebox{-0.08ex}{\greeksym{m}}}m in the experiments shown in
Fig.\,\ref{TimeChart}.

% The quantity $R$ is the ratio of the mass currents flow through the aperture
% and the parallel channel in the absence of trapped currents.

\subsection{Phase-slippage experimental results}

The observation of phase slips in $^4$He has led to a number of quite
significant results that have shed light onto the previously indecipherable
critical velocity problem. I summarise below the most important qualitative
features and their implications.  
\begin{enumerate}
\item{ The critical velocity threshold, which can be seen on time charts such
    as that shown in Fig.\,\ref{TimeChart}, is markedly temperature-dependent
    down to below 200 mK and reaches a well-defined plateau below 150 mK.
    These features can been seen in Fig.\,\ref{vc} and will be discussed
    below. As the thermodynamic properties of superfluid $^4$He are very
    nearly independent of temperature below 1 K, this observation indicates
    that the critical process in action is not governed solely by
    hydrodynamics. It can be suspected that mechanical statistics plays a
    leading role.  }
\item{ Aperture-size is
    not found to be a relevant factor. This feature and the temperature dependence
    mentioned above are in sharp contrast with the Feynman critical velocity,
    which, according to Eq.(\ref{FeynmanVelocity}), exhibits a
    well-characterised dependence on size and none on temperature.}
\item{ The actual velocity threshold for phase slips shows significant scatter
    from one slip to the next in a given sequence, as can be seen in
    Fig.\,\ref{TimeChart}.  This scatter lies much above the background noise
    level of detection of the peak amplitudes of the resonator motion. It
    represents a genuine stochastic property of the process at work, which
    turns out to display a temperature dependence similar to that of the
    critical velocity shown in Fig.\,\ref{Delta-vc}.  }
\item{ The phase slip pattern shows quite reproducible properties in the
    course of a given cool-down as long as the experimental cell is kept at a
    temperature below 10$\sim$15 K. If the temperature is cycled up to
    nitrogen temperature or above, small changes to the critical threshold and
    the pattern itself can occur. This is likely due to changes in the surface
    state of the cell, {\it i.e.} contamination of the micro-aperture walls by
    solidified gases. }
\item{ Quite importantly, phase slips are the signature that {\it quantised
      vortices} are created in aperture flow above a well-defined threshold of
    flow velocity.  This statement arises from the highly reproducible phase
    change, which is measured to be very nearly $2\pi$ and to amount to
    changes of precisely one quantum of circulation in the superfluid loop
    threading the micro-aperture and the long parallel channel (see
    Fig.\,\ref{Cell}). A detailed scenario for the occurence and development of
    phase slips that shows how the phase difference by 2$\pi$ develops has
    been described by Burkhart {\it et al} \cite{Burkhart:94} and is discussed
    below. Different mechanims have been proposed
    \cite{Josserand:95,Josserand:95b,Andreev:04} for which it is unclear that
    the end product of the nucleation process is actually a vortex. }
\end{enumerate}

Critical velocities and phase slips in the superfluid phases of $^3$He show
different features that will be briefly touched upon in Sect.\ref{AllThat}.

\subsection{Several kinds of critical velocities}

\begin{figure}[t]      
  \begin{center}
    \includegraphics[height=7cm]{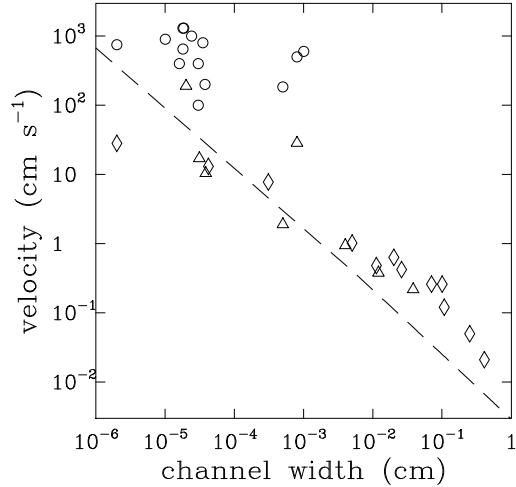}
    \caption{ \label{TwoCriticalVelocities} Critical velocity data {\it vs}
      channel width $\diamond$ -- older data \cite{Wilks:67}; $\circ$ and
      $\scriptstyle\triangle$ -- temperature-dependent and
      temperature-independent data, from \protect{Ref.\cite{Varoquaux:91}}.
      For the temperature-dependent data, the highest value, {\it i.e.} that
      at the lowest temperature, has been retained. The dash-dash line is
      obtained from the Feynman criterion,
      Eq.\protect{(\ref{FeynmanVelocity})}.  }
  \end{center}
\end{figure}

The compilation of the critical velocity data in various apertures and
channels from various sources available in the literature presented at the
Exeter Meeting in 1990 \cite{Varoquaux:91} and shown in
Fig.\,\ref{TwoCriticalVelocities} has not been updated. Two different critical
velocity regimes appear clearly on the graph in
Fig.\,\ref{TwoCriticalVelocities}, a fast regime for small apertures, of the
phase-slip type, and a slower regime for larger channels, of the Feynman type.
More recent data confirm this behaviour. In some occasions, switching between
these two types of critical velocity has been observed in the course of the
same cool-down \cite{Hulin:74,Zimmermann:93}.

% ---------------------------------------------------------------------------
% ---------------------------------------------------------------------------

The data points from various sources \cite{Varoquaux:91} for these two
different types of critical velocity do not fall on well-defined lines as can
be seen in Fig.\,\ref{TwoCriticalVelocities} but merely bunch into clusters of
points. As already stated, critical velocity values in apertures and
capillaries are not very reproducible from experimemt to experiment,
indicating that some less-well-controlled parameters, besides size,
temperature and pressure, also exert an influence.

As a basis for comparison, it is worthwhile to also mention the findings of
the ion propagation studies in superfluid $^4$He at various pressures, which
have been reviewed by McClintock and Bowley
\cite{McClintock:91,McClintock:95}.  Ions can be created in liquid helium and
accelerated by electric field until they reach a critical velocity. The
resulting drift velocities are measured by time-of-flight techniques.  For
negative ions, hollow bubbles 30 \AA\ in diameter with an electron inside, two
different behaviours are observed:
\begin{list}{$\bullet$}{}
\item
Below about 10 bars, vortex rings are
created, on the core of which the electron gets trapped: the drift velocity
suddenly drops from that of the negatively charged bubble to the much slower
vortex velocity \cite{Rayfield:64}.
\item
Above 10 bars, the accelerated ion runs into the roton creation barrier before
vortex rings can be created.  The Landau critical velocity is observed to
decrease from about 60 m/s at SVP down to 46 m/s at 24 bars as the roton
parameters change with pressure while the vortex creation velocity increases
with pressure. 
\item
Around 10 bars, both critical velocities, the Landau critical
velocity for the formation of rotons and that for the formation of vortex
rings can be observed to occur simultaneously because ions can be accelerated
above the threshold for roton emission. 
\end{list}

These ion propagation measurements provide a vivid illustration not only of
the existence of a critical velocity obeying the Landau criterion but also
that roton creation and vortex formation constitute different phenomena and
can exist concurrently \cite{attempt}. The vortex emission threshold displays
other noteworthy features. It depends on temperature in a non-trivial way,
comparable to that of the phase-slip critical velocity with the appearance of
a plateau below $\sim\,$300mK. It also shows the marked dependence on $^3$He
impurity concentration observed for phase slips in micro-aperture flows but
not in larger channels. In both ion propagation and aperture flow
measurements, vortex formation displays very similar features.

Altogether, a careful study of the experimental data in superfluid $^4$He
reveals three different, well-defined, types of critical velocities, one that
is the celebrated Landau critical velocity, another that seems related to the
Feynman criterion with all the uncertainties on the hydrodynamical process of
vortex creation in larger channels, and a third, for phase slips, which is in
want of an explanation: how are the vortices of phase slips in aperture flow
created, and how does the situation differ from that in larger channels?

The short answer, based on qualitative evidence, is that the temperature
dependence of $v_\m c$ and its stochastic properties clearly point toward a
process of nucleation by thermal activation above $\sim$150 mK or so and by
quantum tunnelling below. This conclusion contradicts our daily observations
of the formation of whirlpools and eddies. It will be seen to hold in $^4$He
because the nucleated vortices have nanometric size, a fact that came to be
appreciated because of the detailed analysis of phase slippage observations
that I briefly relate below.

% %%%%%%%%%%%%%%%%%%%%%%%%%%%%%%%%%%%%%%%%%%%%%%%%%%%%%%%%

\section{Phase slip critical velocity} 

% %%%%%%%%%%%%%%%%%%%%%%%%%%%%%%%%%%%%%%%%%%%%%%%%%%%%%%%%

A more firmly established answer to the questions formulated above comes from a
quantitative analysis of the experimental data for phase slips. These
experiments do provide clues that, pieced together, conclusively show that, in
small apertures, vortices are indeed nucleated by thermal activation above
about 150 mK, and by quantum tunelling below.

Let me begin with some preliminary remarks. A glance at Fig.\,\ref{TimeChart}
reveals that the critical velocity threshold itself needs to be defined. Also,
the local value of the critical velocity is not measured directly. Experiments
record the mean value of the volume flow, which is assumed to be proportional
to the local values of the flow field velocity; this assumption breaks down in
the presence of vorticity and has to be taken with a grain of salt (see
Sec.\ref{AllThat}). The value of the critical threshold is not even
reproducible from one cool-down to the next with the same experimental cell.
This lack of reproducibility in the measurements, both in micro-apertures and
in larger channels, has obscured the critical velocity problem for a long
time. It must however be considered as an integral part of the problem.

\begin{figure}[t]
  \begin{center}
    \includegraphics[height=11cm,angle=90]{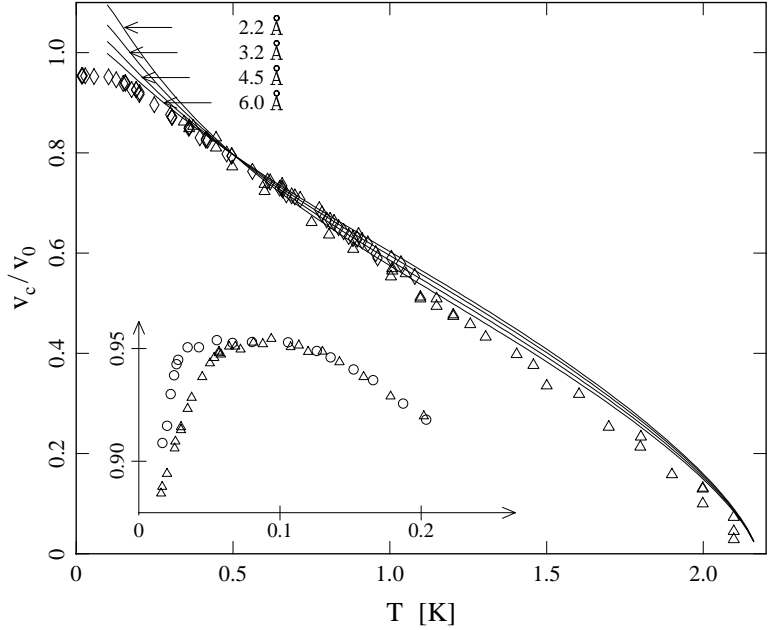}
    \caption{ \label{vc} Critical velocity, normalised to the zero temperature
      {\it linear} extrapolation value $v_0$, {\it vs} $T$, in
      kelvin: ($\diamond$), Ref.\cite{Avenel:93}, for ultra-pure $^4$He;
      ($\scriptstyle\triangle$), Ref.\cite{Zimmermann:98}.  The plain curves
      are computed from the half-ring model (see Sect.\ref{HalfRingModel}) for
      $a_0$ = 2.2, 3.2, 4.5, 6.0 \AA\ and are normalised to match the
      experimental value at 0.5 K. The inset shows the influence of $^3$He
      impurities on $v_\m c$: ($\circ$), 3 ppb $^3$He in $^4$He;
      ($\scriptstyle\triangle$), 45 ppb, from Ref.\cite{Varoquaux:93}.}
  \end{center}
\end{figure}

Now, on with the real topic of this Dossier: nucleation. The first piece of
evidence for the nucleation of vortices, that is their creation {\it ex
  nihilo}, rests on the temperature dependence of the phase-slip critical
velocity shown in Fig.\,\ref{vc}, which increases in a near-linear manner when
the temperature decreases from 2 K to $\sim\,$0.2 K. I mean by linear a
functional dependence going as $v_\m c = v_0(1-T/T_0)$.  As can be seen in
Fig.\,\ref{vc}, the data depart from this linear dependence below 200 mK, where
they reach a plateau, and above 2 K because the critical velocity goes to zero
at $T_\lambda$.

% Normalisation to $v_{\m c0}$. 
% , except in one occasion where it was found by indirect means to peak in the
% vicinity of 21 m/s \cite{Varoquaux:93}
  
% decomposition of velocity field into potential and vortical parts. 
  
This temperature dependence, first observed in 1985 at
Orsay \cite{Varoquaux:87} is now a well-established experimental
fact \cite{Varoquaux:01}. It came as a surprise at first because the critical
velocities observed before were temperature-independent below $\sim$1 K. As
the quantum fluid is nearly fully in its ground state below 1~K -- the normal
fluid fraction becomes less than 1~\% -- one is led to suspect, as was done in
Ref.\cite{Varoquaux:87}, that an Arrhenius-type process must come into play.
If such is the case, that is, if a thermal fluctuation in the fluid with an
energy of at most a few $k_\m B T$ can trigger the appearance of fully-formed
vortex out of nowhere, the energy of this vortex must also be of the order of a few
$k_\m B T$: it must be a very small vortex. But very small vortices require
rather large superfluid velocities to sustain themselves. A careful analysis of
the situation is thus in order.

The nucleation rate for thermally activated process is given in terms of the 
activation energy by Arrhenius'\,law:
\begin{equation}        \label{Kramers}
{\mathit\Gamma}_{\mathrm K} =
\frac{\omega_0}{2\pi}\,\left[(1+\alpha^2)^{1/2}-\alpha\right] 
          \exp\left\{-\frac{E_\m a}{k_{\m B}T}\right\}\;.
\end{equation}
where $\omega_0/2\pi$ is the attempt frequency and $E_\m a$ the activation
energy. The correction for dissipation has been introduced by Kramers to
describe the escape of a particle trapped in a potential well and interacting
with a thermal bath in its environment. The particle undergoes Brownian motion
fluctuations and experiences dissipation. This dissipation is characterised by
a dimensionless coefficient $\alpha = 1/2\omega_0\tau$, $\tau$ being the time
of relaxation of the system toward equilibrium. In superfluid helium,
dissipation is small, although some dissipation is necessary for the system to
reach equilibrium with its environment. Its influence on the thermal
activation rate is very small and will be neglected in the following. However,
this will not be the case anymore in the quantum regime.

Let us derive the expression for the critical velocity that stems from the
Arrhenius rate, Eq.(\ref{Kramers}).  In experiments performed in a Helmholtz
resonator, such as those shown in Fig.\,\ref{TimeChart}, the velocity varies
periodically at the resonance frequency as $v_\m p\,\cos(\omega t)$, $v_\m p$
being the peak velocity. The probability that a phase slip takes place during
the half-cycle $\omega t_\m i = -\pi/2\, ,\; \omega t_\m f = \pi/2$ is
\begin{eqnarray}        \label{probability}
  p &=& 1 - \exp\left\{-\int_{t_\m i}^{t_\m f}
    \mathit\Gamma(P,T,v_\m p\cos(\omega t')\right\} dt'
    \nonumber \\                         
    &=& 1 - \exp\left\{
       -\frac{\omega_0}{2\pi\omega}\sqrt{\frac{-2\pi k_\m B T}
       {v_\m p \left. \left. \partial{E_\m a} \right/ \partial{v} \right|_{\m t=0}}}
      \exp \left\{ -\frac{E_\m a}{k_\m B T} \right\}
      \right\} .     
\end{eqnarray}   
Equation (\ref{probability}) results from an asymptotic evaluation of the
integral at the saddle point $t=0$. The accuracy of the asymptotic evaluation
(\ref{probability}) becomes questionable for $T\rightarrow 0$ as the
energy barrier vanishes. But, as we shall see, quantum effects take over and
the energy barrier never actually vanishes.
%
% ---------------------------------------------------------------------------
% ---------------------------------------------------------------------------

% ---------------------------------------------------------------------------
\begin{figure}[ht]
  \begin{center}
    \includegraphics[height=11cm,angle=90]{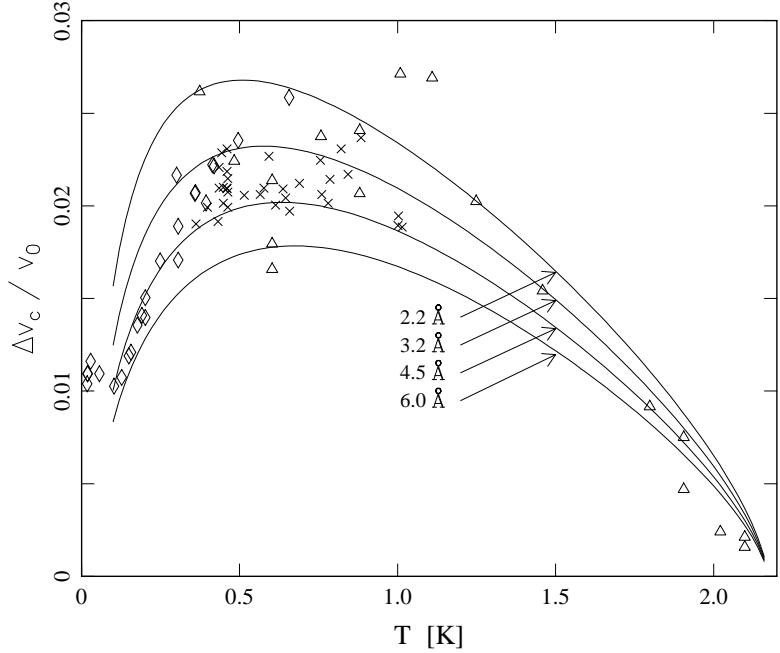}
    \caption{ \label{Delta-vc} 
      Statistical width of the critical velocity
      transition, normalised to the linear extrapolation limit at $T=0$,
      $v_0$, in terms of temperature: ($\diamond$), Ref.\cite{Avenel:93};
      ($\scriptstyle\triangle$), Ref.\cite{Zimmermann:98}; ($\times$),
      Ref.\cite{Steinhauer:95}.} 
  \end{center}
\end{figure}
% ---------------------------------------------------------------------------

The critical velocity $v_\m c$ is defined as the velocity for which $p=1/2$. This
definition is independent of the experimental setup, except for the occurrence in
Eq.(\ref{probability}) of the natural frequency of the Helmholtz resonator
$\omega$. The implicit relation between $v_\m c$ and $E_\m a$ then reads:
\begin{equation}        \label{CriticalVelocity}
\frac{\omega_0}{2\pi\omega}\sqrt{\frac{-2\pi k_\m BT}
       {v_\m c \left. \left. \partial{E_\m a} \right/ \partial{v}
         \right|_{v_\m c}}}
      \exp \left\{ -\frac{E_\m a(P,T,v_\m c)}{k_\m B T} \right\} = \ln 2 \;  .
\end{equation}      
%
% If $E_a$ is given by (\ref{EnergyBarrier}), (\ref{CriticalVelocity}) can be solved
% analytically \cite{Avenel:93}:
% %
% \begin{equation}        \label{Simplev_c}
% v_\m c = v_{\m c0} \left\{1 - \left[\frac{3}{2}\,\frac{k_\m B T}{E_\m J}\gamma 
%          \right]^{\frac{2}{3}} \right\}^{\frac{1}{2}} \;,
% \end{equation}
% %
% where, with logarithmic accuracy, 
% %
% \begin{equation}        \label{gamma}
%   \gamma = \ln{\left\{\frac{\omega_0}{\omega\ln 2}\left[
%   \frac{k_\m B T}{4\pi\,E_\m J\,(1-v^2/v_{\m c0}^2)^{1/2}}
%   \right]^\frac{1}{2}\right\}} 
%   \simeq \ln(0.1 \omega_0/\omega) \; .
% \end{equation}
% %

We note that, in Eq.(\ref{CriticalVelocity}), the attempt frequency is
normalised by the resonator drive frequency: the Brownian particle attempts to
escape from the potential well at rate $\omega_0/2\pi$ but an escape event is
likely only in the time window in a given half-cycle of the resonance during
which the energy barrier stays close to its minimum value $E_\m a(v_\m c)$.
This time interval is inversely proportional to $\omega$, which explains why
an instrumental parameter gets its way into Eqs.(\ref{probability}) and
(\ref{CriticalVelocity}).

The velocity at which individual critical events take place is a stochastic
quantity. Its statistical spread can be characterised by the `width' of the
probability distribution defined \cite{Avenel:93,Zimmermann:90} as the inverse
of the slope of the distribution at $v_\m c$, $\left({\partial p/\partial
    v|_{v_\m c}}\right)^{-1}$. This critical width is found to be expressed
by:
\begin{equation}        \label{Width}
\Delta v_\m c = - \frac{2}{\ln 2}
       \left[
       \frac{1}{2}\left\{
       \frac{1}{v_\m c}+\left.\left. \frac{\partial^2E_\m a}{\partial
         v^2}\right|_{v_\m c}
       \right/\left.\frac{\partial E_\m a}{\partial v}\right|_{v_\m c}
       \right\}
       + \frac{1}{k_\m B T}\left.\frac{\partial E_\m a}{\partial
           v}\right|_{v_\m c}
       \right]^{-1} \; .
\end{equation}

In the experiments, at low temperatures and large critical
velocities, the quantity in curly brackets in the right hand side of
Eq.(\ref{Width}) is small with respect to the last term so that the width is
simply expressed as $\Delta v_\m c = -({2}/{\ln 2})\; k_\m B T
\left(\left.{\partial E_\m a}/{\partial v}\right|_{v_\m c}\right)^{-1}$.
Thus, the statistical width is an approximate measure of the inverse of the
slope of $E_\m a$ in terms of $v$.

\begin{figure}[t]
  \begin{center}
    \mbox{\hskip -3 mm
    \includegraphics[width=72mm]{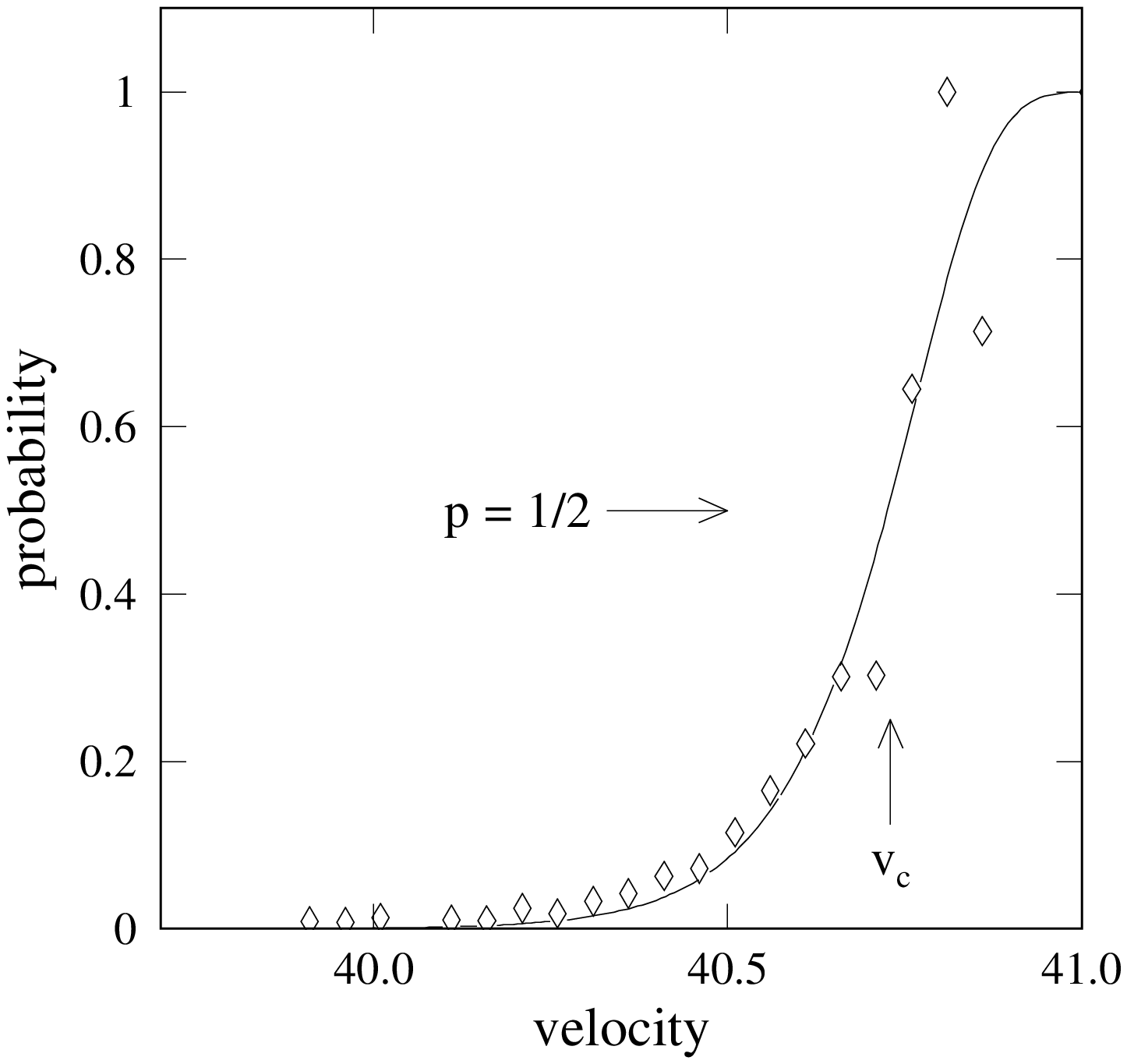}
    \includegraphics[width=72mm]{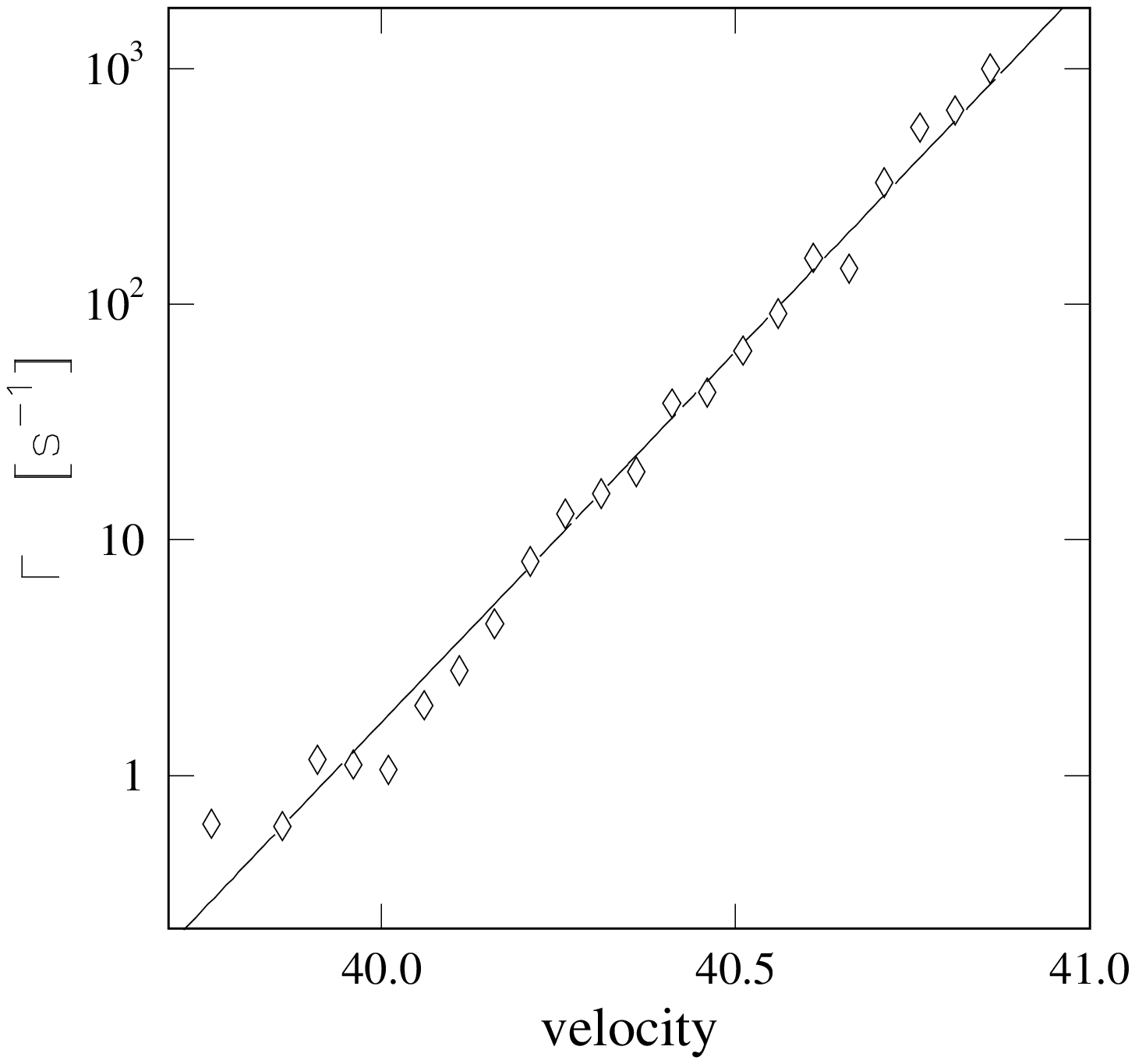}
    }
    \caption{ \label{PR02-pure} Left: Probability $p$ {\it vs} slip velocity
      in winding number.  The plain curve is a non-linear least square fit to
      the analytical form (\ref{probability}), which contains two unknown
      parameters, $v_\m c$ and $\Delta v_\m c$. The critical velocity is
      defined as that for which $p=1/2$.
      \newline 
      Right: Nucleation rate $\mathit\Gamma$ expressed in s$^{-1}$ {\it vs}
      slip velocity in winding number in ultra-pure $^4$He at 17.70 mK and
      saturated vapour pressure on a semi-logarithmic scale. The line is a
      linear fit to the data. }
  \end{center} 
\end{figure}

We now have precise definitions for $v_\m c$ and $\Delta v_\m c$.  These
quantities are derived from $p$, itself obtained by integrating the histograms
of the number of nucleation events ordered in velocity bins. The outcome of
this procedure is illustrated in Fig.\,\ref{PR02-pure}: $p$ shows an
asymmetric-{\sl S} shape characteristic of the double exponential dependence
of $p$ on $v$, Eq.(\ref{probability}), a consequence of Arrhenius'\,law,
Eq.(\ref{Kramers}), being plugged into a Poisson probability distribution. The
observation of this asymmetric-{\sl S} probability distribution constitutes
another experimental clue for the existence of a nucleation process.

The quantities $v_\m c$ and $\Delta v_\m c$ are easily extracted from the
probability curves $p(v)$, but going from $v_\m c$ and $\Delta v_\m c =
-({2}/{\ln 2})\; k_\m B T \left(\left.{\partial E_\m a}/{\partial
      v}\right|_{v_\m c}\right)^{-1}$ back to $E_\m a(v)$ and $\omega$ by
numerical integration of the differential equation (\ref{CriticalVelocity})
requires more work and introduces additional errors.  As discussed in
Ref.\cite{Varoquaux:03}, an improved procedure consists in obtaining directly
the escape rate $\mathit\Gamma$ from the phase slip data.  This quantity is
the ratio, for a given velocity bin, of the number of slips which have
occurred at that velocity to the total time spent by the system at that given
velocity.  The outcome of this procedure is illustrated in
Fig.\,\ref{PR02-pure}.  The slope of $\ln \mathit\Gamma(v)$ directly yields
$\left.{\partial E_\m a/\partial v} \right|_{v_\m c}$; the value of $\ln
\mathit\Gamma$ at $v_\m c$ gives a combination of $\ln\omega_0$ and $E_\m
a(v_\m c)$, which is still not easy to cleanly disentangle
\cite{Varoquaux:86}.

But the experiment itself offers help \cite{Nature} as I now describe.

% ---------------------------------------------------------------------------
% ---------------------------------------------------------------------------

% %%%%%%%%%%%%%%%%%%%%%%%%%%%%%%%%%%%%%%%%%%%%%% 
% 
\section{Vortex nucleation: thermal {\bf{\textit{vs}}} quantum } 
% 
%%%%%%%%%%%%%%%%%%%%%%%%%%%%%%%%%%%%%%%%%%%%%%% 

Below 0.15 K, $v_\m c$ ceases abruptly to vary with $T$, as seen in
Fig.\,\ref{vc}. For ultra-pure $^4$He (less than 1 part in 10$^9$ of
$^3$He impurities), $v_\m c(T)$ remains flat down to the lowest temperatures
($\sim\,$12 mK) reached in the experiment.  The crossover from one regime to
the other is very sharp. At the same crossover temperature $T_\m q$, $\Delta
v_\m c$ also levels off sharply. It is believed on experimental grounds that
this saturation is intrinsic and is not due to stray heating or parasitic
mechanical vibrations; this question is of paramount importance and
considerable efforts have been devoted to lift all uncertainties and
completely elucidate the matter \cite{Avenel:93}.

Even if all possibilities of an experimental artifact are cleared out, the
mere observation of a plateau in $v_\m c$ is no sufficient proof for a
crossover from the thermal regime to the quantum one: the effect of $^3$He
impurities, shown in the insert of Fig.\,\ref{vc}, also gives a levelling-off of
$v_\m c$ {\it vs} $T$. This effect has been studied in detail in
Ref.\cite{Varoquaux:93} and is well understood. Incidentally, it shows that
the phase slip phenomenon taking place in the micro-aperture tracks the
temperature down to below $\sim$ 12 mK, the lowest temperature in these
experiments: there is no spurious temperature saturation effect.

If the nucleation barrier were undergoing an abrupt change at $T_\m q$, for
instance because of a bifurcation toward a vortex instability of a different
nature \cite{Josserand:95}, in all likelihood $\Delta v_\m c$ would jump to a
different value characteristic of the new process (presumably small since
$v_\m c$ reaches a plateau). Such a jump is not observed in
Fig.\,\ref{Delta-vc}. Furthermore, $v_\m c$ levels off below $T_\m q$, which
would imply through Eq.(\ref{CriticalVelocity}) that $E_a$ becomes a very
steep function of $v$, but $\Delta v_\m c$ also levels off, which, through
Eq.(\ref{Width}), would imply the contrary. This remark leads us to
investigate the possibility that, below $T_\m q$, thermally-assisted escape
over the barrier gives way to quantum tunnelling under the barrier
\cite{Ihas:92}. This transition would induce plateaus below $T_\m q$ for both
$v_\m c$ and $\Delta v_\m c$.

I state again that, during the course of these investigations at Orsay-Saclay,
the group of Peter McClintock at Lancaster concluded in their ion
propagation studies to the existence of a crossover around 300 mK from a
thermal to a quantum regime for the nucleation of vortices \cite{Hendry:88},
as predicted by Muirhead, Vinen, and Donnelly \cite{Muirhead:84}. There
certainly are significant differences between the ion limiting drift velocity
and aperture critical flow -- in particular, the latter is nearly one order of
magnitude smaller -- but the qualitative similarities are striking. We thus
have two completely different types of experiments that point toward vortex
nucleation, both in a thermal regime and in a quantum one.

\subsection{The macroscopic quantum tunnelling rate}

To proceed with our investigation, let us now make the assumption that below
$T_\m q$, zero point fluctuations do take over thermal fluctuations. The
potential barrier is not surmounted with the assistance of a large thermal
fluctuation, it is tunnelled under quantum-mechanically; non-conservation
of energy is not a problem if it is brief enough, as stated by the Heisenberg
uncertainty principle for energy. The quantum-tunnelling event is ``assisted''
by the zero point fluctuations \cite{Martinis:88}. What is remarkable here,
and not necessarily easy to admit, is that such an energy non-conserving
process does affect a macroscopic number of atoms, that necessary to form a
vortex of about 50 \AA\ in length, as we shall see below.

Such ``macroscopic quantum tunnelling'' (MQT) processes have been the object
of numerous experimental and theoretical studies, mainly in superconducting
Josephson devices. The case of vortices in helium can be treated in a very
similar manner, as done in Ref.\cite{Varoquaux:03}. Before giving a brief
relation of MQT for vortices in $^4$He, I summarise some of the basic results of
the extended body of theoretical studies that followed Caldeira and
Leggett's original work \cite{Caldeira:83a}.

The quantum tunnelling rate of escape out of a potential well $V(q)$ is a
textbook problem \cite{Landau:barrier}. The rate is proportional to
$\exp{-S/\hbar}$, $S$ being, in the WKB approximation, the action of the
escaping particle along the saddle-point trajectory at the top of the
potential barrier, the so-called ``bounce'' \cite{Coleman:77}. For a particle
of mass $m$ and energy $E$ escaping from a one-dimensional barrier $V(q)$, this
action reads
\begin{equation}        \label{Action}
  S = 2 \int_{q_1}^{q_2} \m d q \sqrt{2m[V(q)-E]} \; .
\end{equation}
The determination of the bounce yields the points $q_1$ and $q_2$ at which
the particle enters and leaves the barrier. 

A discussion of the quantum tunnelling of vortices thus requires a Lagrangian
formulation of vortex dynamics. Such a formulation has been carried out in
particular by Sonin \cite{Sonin:95} (see also Ref.\cite{Fischer:00} for an
extended discussion). However, analytical results can be obtained only at the
cost of approximations and yield less than fair comparison with experiments
(see the discussion in Ref.\cite{Varoquaux:01}).

Here, I follow, as in Ref.\cite{Varoquaux:03}, the usual approach taken in the
literature for Josephson devices \cite{Caldeira:83a,Larkin:84}, which is to
choose for $V(q)$ a simple analytic form limited to a parabolic and cubic term
in $q$: 
\begin{equation}        \label{CubicParabolic}
  V(q) = V_0 + \frac{1}{2}m\omega_0^2 q^2 
    \big(1-\frac{2q}{3q_\m b}\big)\; ,
\end{equation}
where $\omega_0$ is the angular frequency of the lowest mode of the trapped
particle and $q_\m b$ the generalised coordinate of the barrier top location. The
barrier height $E_\m b$ is equal to $m\omega_0^2 q_\m b^2/6$.

This simple form is of general applicability when the applied velocity is
close to the limit, which I call $v_{\m c0}$, where the energy barrier
vanishes and the system ``runs away'', the so-called ``lability'' point.  At
this point, the critical velocity is reached even in the absence of thermal or
quantum fluctuations.  Such a hydrodynamic instability threshold at which
vortices appear spontaneously has been shown to occur in numerical simulations
of flows past an obstacle using the Gross-Pitaevskii equation by Frisch et al.
\cite{Frisch:92} and others \cite{Nore:00,Berloff:01,Rica:01}.

The zero-temperature WKB tunnelling rate for the cubic-plus-parabolic
potential $E_\m b$, Eq.(\ref{CubicParabolic}), is found to be  \cite{Caldeira:83a} 
\begin{equation}        \label{ZeroTemperature}
  \mathit\Gamma_0 =\frac{\omega_0}{2\pi}\, \big(120\pi\, \frac{S_0}{\hbar}\big)^{1/2}
    \exp -\frac{S_0}{\hbar} \; ,   
\end{equation}
the action $S_0$ being equal to $36 E_\m b/5\omega_0$. 

From this result, we may anticipate that the crossover between the quantum and
the thermal regime lies around a temperature close to that for which the
exponents in Eqs.(\ref{Kramers}) and (\ref{ZeroTemperature}) are equal, namely $T =
5\omega_0/36 k_\m B$ -- assuming that the activation energy in
Eq.(\ref{Kramers}), $E_\m a$, reduces to the simple cubic-plus-parabolic form,
$E_\m b$. A more precise study of the mathematical properties of the
quantum channel for escape leads to the following relation 
\cite{Melnikov:91,further}
\begin{equation}        \label{Crossover}
  \hbar \omega_0 = 2 \pi k_\m B T_\m q \; .
\end{equation}
Once the crossover temperature has been determined from experiment, $\omega_0$
is fixed to pinpoint accuracy compared to the fitting procedure outlined in
the previous Section. This is where we get help from the experiment
because both the value of $\omega_0$ is now completely pinned
down and the interpretation of the experiment in terms of a nucleation process
is confirmed.  The values of the barrier height $E_\m b$ at each given
velocity then follow easily, using the full expressions for the rate in terms
in terms of $E_\m b$, $\omega_0$ and, also, the damping parameter $\alpha$.

\subsection{Friction in MQT}

Damping turns out to matter significantly in the quantum tunnelling of
semi-macroscopic objects. The relevance and applicability of the concept of
quantum tunnelling to macroscopic quantities such as the electric current
through a Josephson junction or the flow of superfluid through a
micro-aperture, although still sometimes questioned, have been checked in
detail for the Josephson effect case \cite{Martinis:87}. One of the conceptual
difficulties, besides the large number of particles involved, is that the
macroscopic system is coupled to an environment that acts as a thermal bath;
this coupling gives rise to a source of fluctuations and semi-classical
friction. This issue was tackled by Caldeira and Leggett \cite{Caldeira:83a},
and a number of other authors (see, for instance, \cite{Melnikov:91,further}).
In the case of weak ohmic damping ($\alpha \ll 1$) and for the
cubic-plus-parabolic potential, the tunnelling rate takes
the form \cite{Caldeira:83a,Grabert:87,Waxman:85}:
\begin{equation}        \label{QuantumRate}
  \begin{split}
    {\mathit\Gamma}_{\m{qt}}= 
    & \frac{\omega_0}{2\pi}
    \,\left(864\pi\,\frac{E_\m b}{\hbar\omega_0}\right)^{1/2} 
    \\
    &\times\exp\left\{-\,\frac{36}{5}\,\frac{E_\m b}{\hbar\omega_0}
    \left[1+\frac{45\zeta(3)}{\pi^3}\alpha\right] \right. 
    \left. +\frac{18}{\pi}\alpha\,\frac{T^2}{T_\m q^2}+
    {\mathcal O}\left(\alpha^2,\,\alpha\,\frac{T^4}{T_\m q^4}\right)\right\}  \; .
  \end{split} 
\end{equation}
Thus, according to Eq.(\ref{QuantumRate}), damping depresses the MQT escape
rate at $T=0$ -- $\alpha$ is a positive quantity -- and introduces a temperature
dependence that increases the rate as $T$ increases. These effects are large,
even for weak damping, because they enter the exponent of the exponential
factor in Eq.(\ref{QuantumRate}).  Relation (\ref{Crossover}) between $T_\m q$
and $\omega_0$ is nearly unaffected by damping: $\omega_0$ is simply changed
into $\omega_0[(1+\alpha^2)^{1/2} -\alpha]$ according to Eq.(\ref{Kramers}), a
minor modification for $\alpha\ll 1$.

% ---------------------------------------------------------------------------
\begin{figure}[t]
  \begin{center}
    \includegraphics[width=60 mm,angle=90]{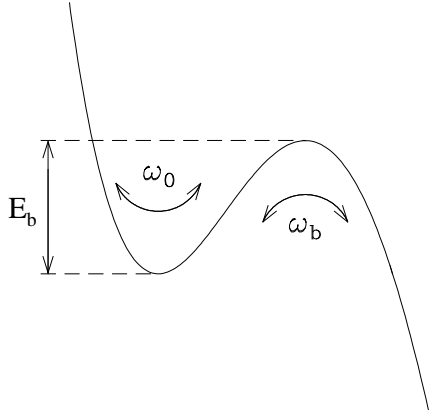}
    \caption{ \label{PotentialWell} Potential well trapping a particle in one
      dimension. The particle can escape to the continuum of states to the
      right. The lowest mode at the bottom of the well has angular frequency
      $\omega_0$; $\omega_\m b$ would be the corresponding quantity if the
      potential was inverted bottom over top. There can exist intermediate
      energy levels within the well, which are populated according to the
      Boltzmann factor. }
  \end{center}
\end{figure}
% ---------------------------------------------------------------------------

Equation (\ref{QuantumRate}) is valid up to about $T_\m q/2$. From $T_\m
q/2$ to $\sim T_\m q$, one has to resort to numerical calculations
\cite{Grabert:87}. In the thermal activation regime, $T\gtrsim T_\m q$,
quantum corrections affect the Kramers escape rate up to about $3\,T_\m q$ and
can be evaluated analytically. These high-temperature quantum corrections
depend only weakly on friction.  A complete solution of the problem of the
influence of friction, weak, moderate or strong, in the regime where thermal
fluctuations still prevail but quantum corrections cannot be neglected has
first been worked out in the classical regime ($T\gg T_\m q$) by Grabert
\cite{Grabert:88}. The extension to the temperature range $T\gtrsim T_\m q$
was then carried out by Rips and Pollak \cite{Rips:89} who showed that the
rate for arbitrary damping in the temperature range $T>T_\m q$ can be
factorised into physically meaningful terms:
\begin{equation}         \label{GeneralKramers}
{\mathit\Gamma} = \,f_\m q\,{\mathit\Upsilon}\,\mathit\Gamma_\m K \;,
\end{equation}
namely, the classical Kramers rate ${\mathit\Gamma}_\m K$, the quantum correction
factor $f_\m q$, and the depopulation factor $\mathit\Upsilon$. The high
temperature limit of $f_\m q$ is
\begin{equation}             \label{LeadingQuantumCorrection}
f_\m q  =
\exp \left\{\frac{\hbar^2}{24} \frac{(\omega_0^2+\omega_\m b^2)}{(k_{\m B}T)^2}
     +{\mathcal O}(\alpha/T^3,1/T^4)\right\}  \; ,
\end{equation}
in which $\omega_0$ and $\omega_\m b$ are the confining potential parameters
depicted in Fig.\,\ref{PotentialWell}. Analytic results for $f_\m q$ are known
to slightly below $T_\m q$ \cite{Grabert:87,Hanggi:90}.

The depopulation factor $\mathit\Upsilon$ describes the fact that the escape
process eventually depletes the occupancy of the energy levels inside the
potential well. This factor, the expression of which is too bulky to be
reproduced here (see Ref.\cite{Rips:89}), is unity at large $\alpha$ when the
coupling of the Brownian particle with the thermal bath is large. It decreases
to zero as $\alpha \rightarrow 0$ and the system becomes effectively decoupled
from the environment. In the quantum regime, dominated by zero point
fluctuations, level depletion does not take place and $\mathit\Upsilon$ is
unity. For the nucleation of vortices, friction turns out to always be both
sufficient and not too large so that depopulation corrections remain small and
$\mathit\Upsilon \sim 1$.

% ---------------------------------------------------------------------------
\begin{figure}[t]
  \begin{center}
    \includegraphics[width=70mm]{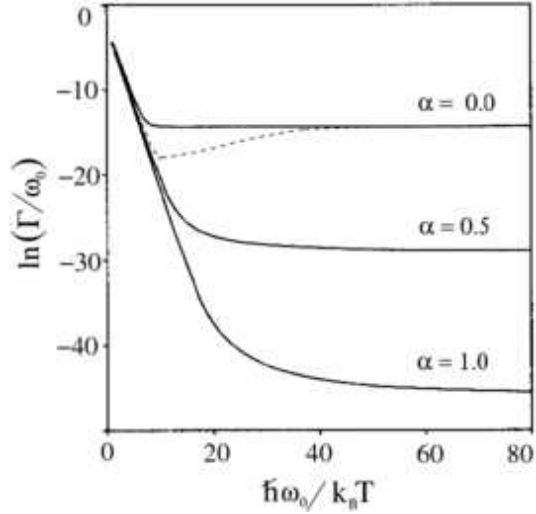}
    \caption{ \label{EscapeRate}Logarithm of the escape rate normalised to the
      attempt frequency in terms of inverse temperature, also normalised to
      $\omega_0$ for various value of the damping parameter, adapted from
      Ref.\protect{\cite{Grabert:87}}. The dot-dot line is a hand-sketch of
      the situation where $\alpha$ increases with temperature, starting from
      zero at $T=0$. }
  \end{center} 
\end{figure}
% ---------------------------------------------------------------------------
 
The escape rate calculated for three values of the damping parameter $\alpha$ over
the full temperature range is shown in Fig.\,\ref{EscapeRate}. A hand sketch
shows the influence of a temperature dependence in the damping
coefficient. Such a situation is found in the nucleation of vortices in $^4$He
as I now describe.

\subsection{Experimental energy barrier and damping coefficient}

\begin{figure}[t]
  \begin{center}
    \includegraphics[width=80mm]{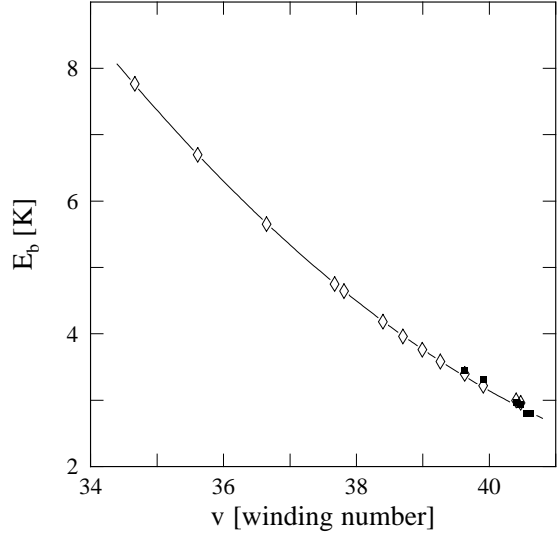}
    \caption{ \label{ExperimentalEb} The barrier energy $E_\m b$ in kelvin
      {\it vs} $v$, the velocity in the aperture expressed in phase winding
      number obtained: ($\scriptscriptstyle{\blacksquare}$), from the LT data
      transformed using the numerical tables in Ref.\cite{Grabert:87};
      ($\scriptstyle\lozenge$), from the high temperature data. The high $T$
      and low $T$ analyses yield consistent results in the region where they
      overlap.}
  \end{center} 
\end{figure}

From this knowledge of the theoretical analytical and numerical expressions
for the rate $\mathit\Gamma$, obtained for the cubic-plus-parabolic potential,
it becomes possible to extract from the measured nucleation rate and crossover
temperature the values of the energy barrier in terms of $v_\m c$.  The value
of $\omega_0$ given by Eq.(\ref{Crossover}) ($\omega_0/2\pi = 2\times10^{10}$
Hz for $T_\m q$=0.147 K) is consistent with the attempt frequency appropriate to
the thermally-activated regime \cite{Varoquaux:86} and that found directly
from the fits to the probability $p$ as shown in Fig.\,\ref{PR02-pure}.
Furthermore, it agrees well (for $a_0$ = 4.5 \AA) with the eigenfrequency of the
highest Kelvin mode that a vortex filament in $^4$He can sustain, $\omega_+ =
\kappa_4/\pi a_0^2 = \omega_0$.  The final step consists to extract the values
for the energy barrier $E_\m b$ from the measured escape rate. These values of
$E_\m b$ in the case of the experiments on ultra-pure $^4$He analysed in
Ref.\cite{Varoquaux:03} are shown in Fig.\,\ref{ExperimentalEb}.

% ---------------------------------------------------------------------------
% ---------------------------------------------------------------------------

The self-consistency of the procedure can be checked by using the values of
$\omega_0$ and $E_\m b$ derived from this analysis of the nucleation rate to
compute $v_\m c$ and $\Delta v_\m c$ using Eqs.(\ref{CriticalVelocity}) and
(\ref{Width}), {\it mutatis mutandis}, and compare with the experimentally
determined values.  We thus have a form of closure procedure to check the
analysis, from which we conclude that our assumption according to which
vortices are nucleated by quantum tunnelling below $T_\m q$ shows full
consistency with the thermally-assisted nucleation regime that prevails above
$T_\m q$.

% ---------------------------------------------------------------------------
\begin{figure}[t]
  \begin{center}
    \includegraphics[width=90mm]{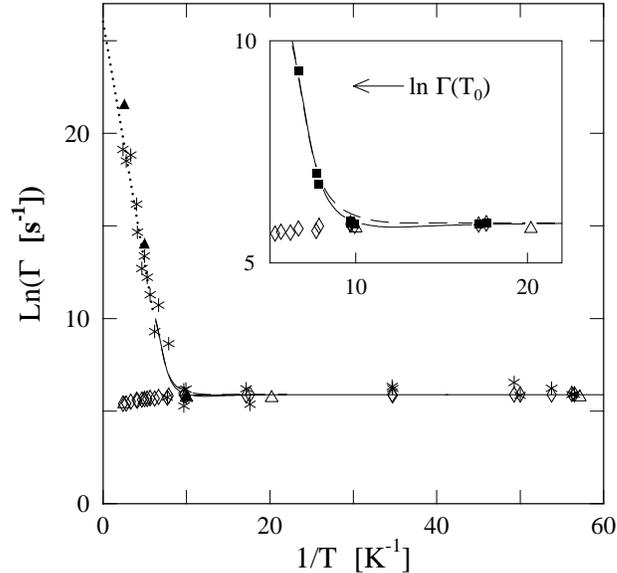}
    \caption{ \label{ArrheniusPlot} $\ln \mathit\Gamma(v)$ {\it vs} $1/T$,
      $\mathit\Gamma$ being expressed in s$^{-1}$ and $T$ in K:
      ($\scriptstyle\lozenge$), as measured at varying $T$ and $v_\m c$;
      ($\ast$), corrected for the change of the velocity with $T$ as explained
      in the text. The raw data from the run with the second sample of
      ultra-pure $^4$He, ($\scriptstyle\triangle$),
      ($\blacktriangle$), agree very well with that of the first
      sample.
      \newline
      In the inset, $\ln \mathit\Gamma(v_\m q)$,
      ($\scriptscriptstyle\blacksquare$), has been obtained with smoothed
      values of $v_\m c$.  The curves represent the calculated values of $\ln
      {\mathit \Gamma(v_\m q)}$ with $\alpha= 0$ (dash-dash) or varying with
      $T$ (plain) as explained in the text.  The dot-dot curve is the
      extrapolation to $1/T=0$ of a linear fit to the high temperature portion
      of the data.  }
  \end{center} 
\end{figure}
% ---------------------------------------------------------------------------

The quantitative analysis can be carried out one step further by constructing
a Arrhenius plot from the experimental data and comparing directly the outcome
to the results from theory. Arrhenius plots are drawn at constant $E_\m b$ and
varying temperature. Our results here are obtained at velocities that vary
with temperature, hence at varying $E_\m b$'s. As can be noted in
Fig.\,\ref{ArrheniusPlot}, the raw experimental, velocity-dependent, rates
exhibit little variation over the range of parameters: escape rates are only
observed in a certain window determined by experimental techniques. At low
temperatures, $T< T_\m q$, the critical velocity is close to its zero
temperature limit $v_\m q$ and the corrections to $\mathit\Gamma$ are small.
As $T$ increases above $T_\m q$, $v_\m c$ decreases and $\mathit\Gamma$ has to
be determined by piecewise integration of $\m d \ln\mathit\Gamma / \m d v$.
The high temperature extrapolation for $\mathit\Gamma$ obtained in such a
manner does display the expected $1/T$ dependence, as seen in
Fig.\,\ref{ArrheniusPlot}.

The low temperature corrected $\mathit\Gamma$ shows, as can be seen in the
inset of Fig.\,\ref{ArrheniusPlot}, a small, but real, drop below its zero
temperature limit as the temperature is raised, thereby simply following the
trend of the measured $\mathit\Gamma$. As illustrated in Fig.\,\ref{EscapeRate},
this drop reveals the influence of damping. A damping coefficient $\alpha$
that increases from 0 at $T=0$ to $\sim\,0.1$ around $T_\m q$ and more slowly
above accounts for the observed drop \cite{Varoquaux:03}. This $T$-dependent
dissipation also makes the crossover between the thermal and the quantum
regimes even sharper than for $\alpha = 0$, and closer to observations. The
nucleation of vortices in $^4$He thus offers a rare observation of the effect
of damping on MQT.

% %%%%%%%%%%%%%%%%%%%%%%%%%%%%%%%%%%%%%%%%%%%%%%%
 
\section{The vortex half-ring model}\ 
                        \label{HalfRingModel}

% %%%%%%%%%%%%%%%%%%%%%%%%%%%%%%%%%%%%%%%%%%%%%%% 

As described in the previous Section, there is strong evidence that the
experimental features of the phase slip data result from quantised vortex
nucleation. The nucleation barrier $E_\m b$ is of the order of a few kelvins
(see Fig.\,\ref{ExperimentalEb}). The attempt frequency $\sim 2\times10^{10}$ Hz
is of the order of the highest Kelvin waves mode.  In this Section, I wish to
describe a simple model that will account for the features described above.
This model, the nucleation of vortex half-rings at a prominent asperity on the
walls, finds its roots in the work of Langer, Fischer, and Reppy
\cite{Langer:70,Langer:67b}, Volovik \cite{Volovik:72}, and Muirhead, Vinen
and Donnelly \cite{Muirhead:84}. It was further developed and put on firm
experimental findings in Ref.\cite{Avenel:93}.

The model premise is quite simple. Consider, as done by Langer and
Reppy in Ref.\cite{Langer:70}, the homogeneous nucleation of a vortex ring in
a homogeneous flow $v_\m s$. When the ring has reached radius $R$ in a plane
perpendicular to the flow, its energy in the laboratory frame, where the
observer is a rest and sees the superfluid moving at velocity $v_\m s$, is
expressed by
\begin{equation}        \label{VortexFreeEnergy}
  E_\m v = E_\m R - P_\m R v_\m s \; .
\end{equation} 
The rest energy $E_\m R$ and impulse $P_\m R$ of the vortex ring are given by
Eqs.(\ref{RingEnergy}) and (\ref{RingImpulse}). The minus sign in the right
hand side of Eq.(\ref{VortexFreeEnergy}) arises because I have implicitly
assumed that the vortex opposes the flow, {\it i.e.}, that its impulse $P_\m
R$ points straight against $v_\m s$: this configuration minimises $E_\m v$.

\begin{figure}[t]
  \begin{center}
    \includegraphics[width=70mm]{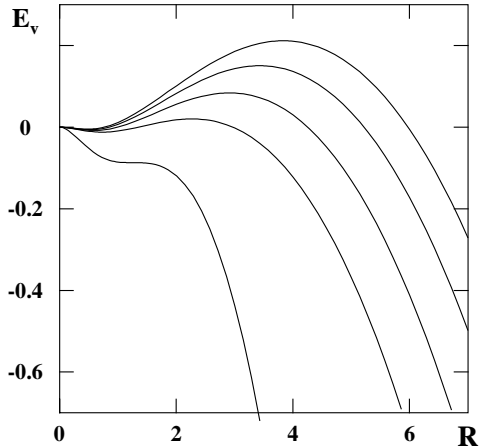}
    \caption{ \label{Ev}Energy barrier $E_\m v$ {\it vs} $R$, the vortex radius,
      in the simple half-ring model at various superfluid flow velocities.
      This figure is taken from Ref.\cite{Avenel:93} and applies for the model
      and with the units used in that reference. It is given here for
      illustration purposes, namely that potentials of the form of
      Eq.(\ref{VortexFreeEnergy}) give rise to confining wells. }
  \end{center} 
\end{figure}

The rest energy $E_\m R$ increases with vortex size as $R\ln R$ and the
impulse $P_\m R$ as $R^2$: the impulse term becomes dominant at large radii
and causes $E_\m v$ to become negative. The variation of $E_\m v$ in terms of
$R$ has the shape of a confining well potential, which becomes shallower and
shallower with increasing $v_\m s$, as depicted in Fig.\,\ref{Ev}. The
barrier height can easily be computed numerically and substituted into the
expression for $v_\m c$, Eq.(\ref{CriticalVelocity}).  An analytical
approximation for $v_\m c$ involving the neglect of logarithmic terms and
valid for large vortices ($R\gg a_0$) has been given by Langer and Reppy
\cite{Langer:70}.

Such a critical velocity would be for the formation of a mist of vortices in
the bulk of the superfluid. However, this sort of vorticity condensation does
not take place for two reasons. Firstly, the velocity of potential flows,
which follows from the Laplace equation, reaches its maximum value at the
boundaries, not in the bulk. Secondly, the nucleation of a vortex half-ring at
the boundary itself involves a half of the energy given by
Eq.(\ref{VortexFreeEnergy}). Hence, half-ring nucleation at the wall is always
much more probable at the same velocity $v_\m s$ than full ring nucleation in
the bulk. Half of the energy for the half-ring holds for classical
hydrodynamics, the other half being taken care of by the image in the plane
boundary. For a superfluid vortex, the actual energy of a half-ring is smaller
than in the classical ideal fluid because the superfluid density is depleted
at the solid wall and the core radius increases. This effect strengthens the
case for half-rings, as discussed in Ref.\cite{Burkhart:94}.

% ---------------------------------------------------------------------------
% ---------------------------------------------------------------------------

The barrier height can easily be computed and substituted into the expressions
for $v_\m c$ and $\Delta v_\m c$, Eqs.(\ref{CriticalVelocity}) and
(\ref{Width}).  Critical velocities $v_\m c$ and statistical widths $\Delta
v_\m c$ computed in such a manner are shown as a function of temperature in
Figs.\ref{vc} and \ref{Delta-vc} for several values of the vortex core
parameter $a_0$.  A value of 4.5 \AA\ gives near-quantitative agreement with
the experimental observations over the entire temperature range.  This value
of $a_0$ is compatible not only with the temperature variations of $v_\m c$
and $\Delta v_\m c$ but also with the magnitude of the {\it local} $v_\m c$
found to be $20\sim22$ m/s using $^3$He impurities as a local velocity probe
\cite{Varoquaux:93}. It exceeds that in the bulk ($a_0\simeq 2.5$ \AA), which
is thought to reflect the proximity of the wall as discussed in greater
details in Ref.\cite{Varoquaux:01}. With this value, the nucleating half-ring
has a radius of approximately 15 \AA\ at the top of the barrier.

Once nucleated, the vortex floats away, carried out by the superfluid stream
at the local superfluid velocity and by its own velocity, $v_\m R = \partial
E_\m R/\partial P_\m R$. It can be noted that, at the top of the barrier, $\partial E_\m v/\partial R =0$ and the vortex self-velocity $v_\m R$ exactly balances the
applied $v_\m s$: the nucleating vortex is at a near standstill. 

\begin{figure}[t]
  \begin{center}
    \mbox{\hskip 3mm\includegraphics[width=53mm,angle=90]{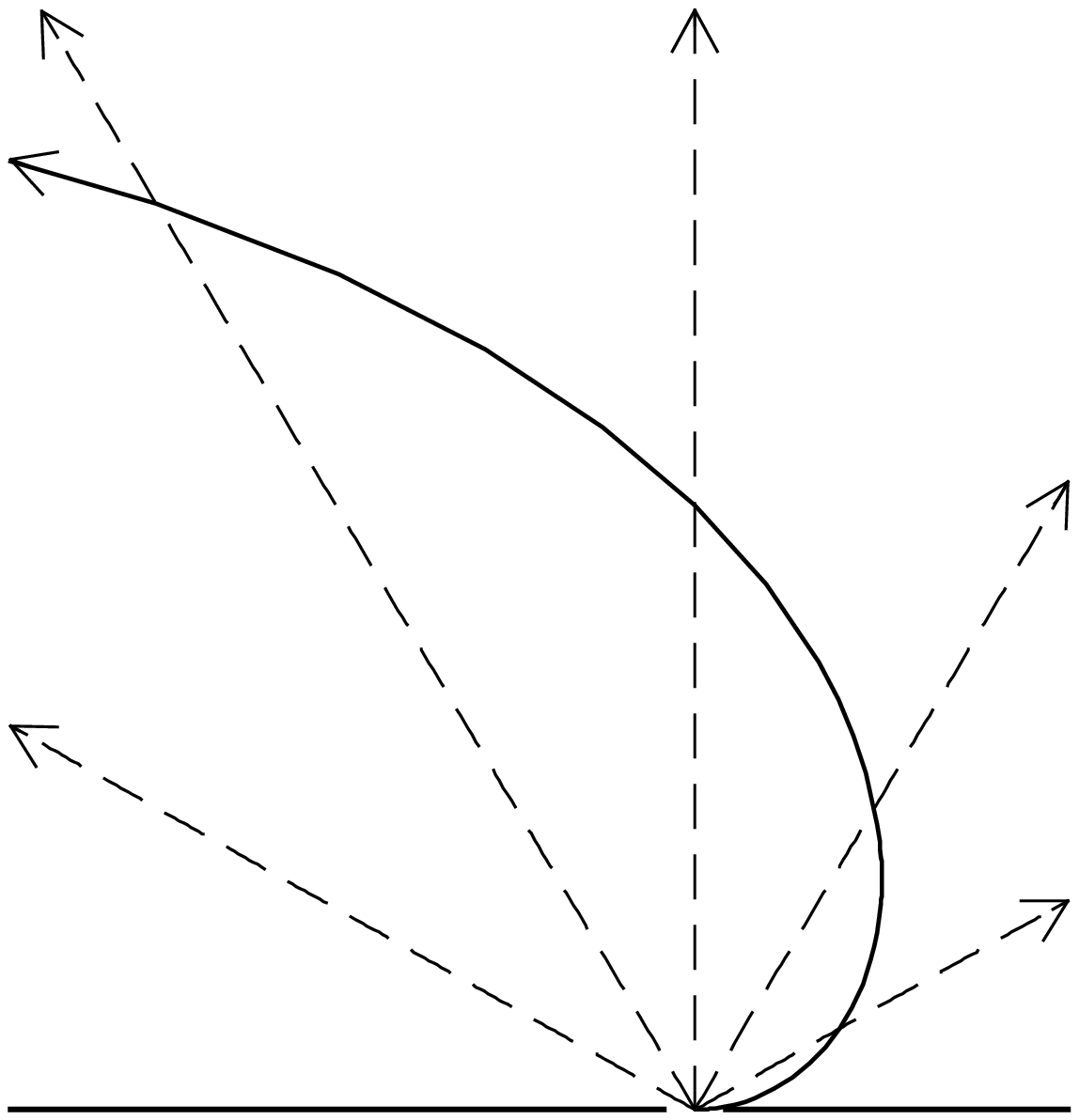}
    \hskip 10 mm\includegraphics[width=70mm]{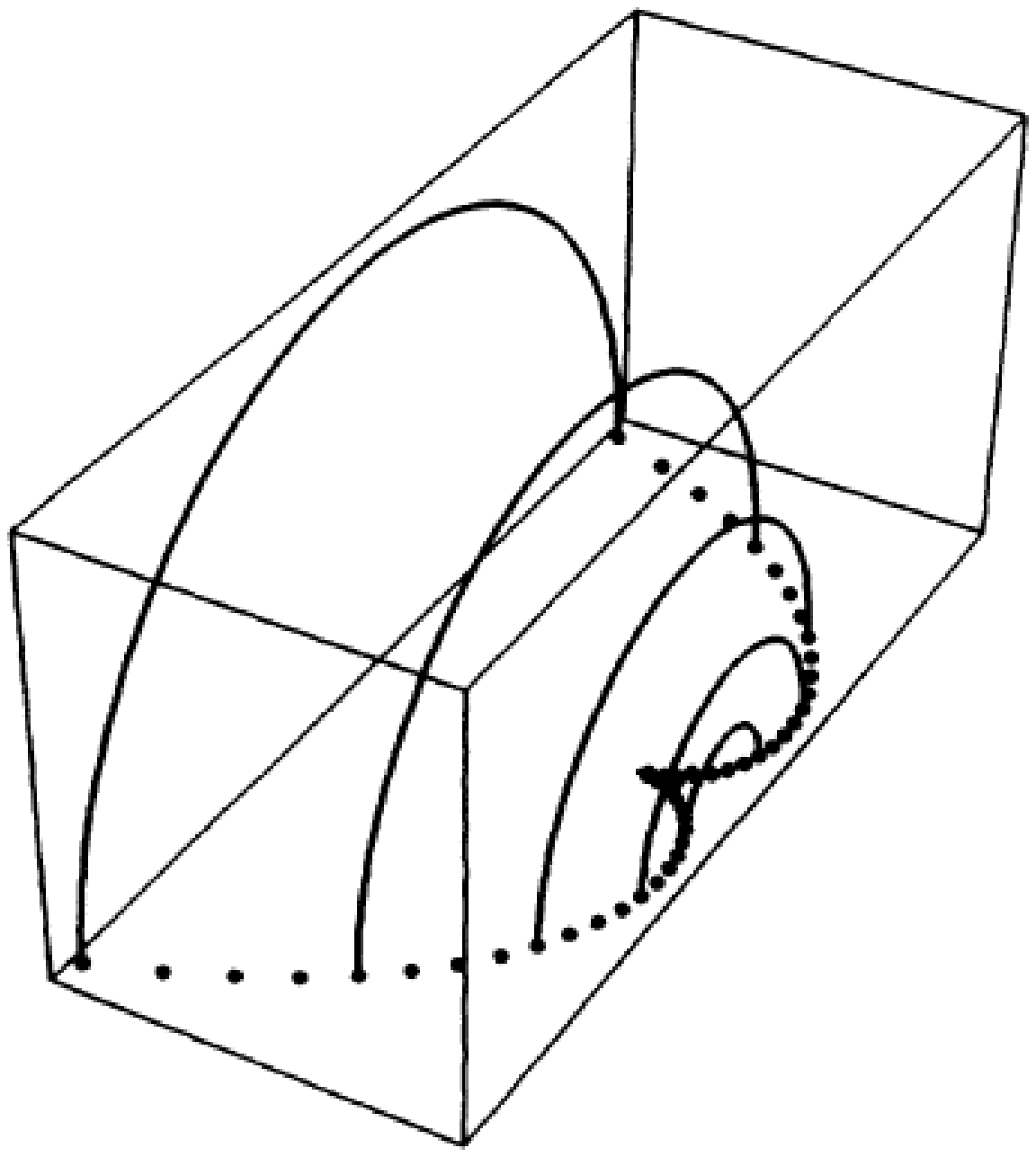}}
  \caption{ \label{trajectory} 2D (left) and 3D (right) views of the vortex
    half-ring trajectory over a point-like orifice in an infinite plane. The
    dash-dash lines on the 2D plot are the potential flow lines that emerge
    from the orifice.}
  \end{center} 
\end{figure}

If the flow is uniform, with parallel streamlines, nothing much happens; the
vortex wanders away and the interaction with the normal fluid and with the
wall causes a loss of vortex energy that eventually leads to its
disappearance. If the flow is divergent, as in Fig.\,\ref{trajectory}, the
vortex tends to follow the local streamlines and grow under the combined
action of the potential flow and its own self-velocity: it then gains energy
at the expense of the potential flow.  In such a way, it can expand from
nanometric to micrometric sizes and above.  The vortex in its motion away from
the micro-aperture takes a finite lump of energy to remote places of the cell.
This energy loss reduces the Helmholtz resonance amplitude in a fairly sudden
manner. Such a dissipative event gives the signature of single phase slips
that is seen in Fig.\,\ref{TimeChart}.

This scenario for a phase slip involves a change of the phase difference
between the two sides of the micro-aperture of exactly $2\pi$ because the
vortex ends up crossing all the streamlines, as pictured in
Fig.\,\ref{trajectory}. This crossing causes the velocity circulation to change
by exactly one quantum $\kappa_4$ on all the superfluid paths extending from
one side of the aperture to the other.

% ---------------------------------------------------------------------------
% ---------------------------------------------------------------------------

% %%%%%%%%%%%%%%%%%%%%%%%%%%%%%%%%%%%%%%%%%%%%%% 
% 
\section{Pinning, vortex mills, collapses and all that}
                        \label{AllThat} 
% 
%%%%%%%%%%%%%%%%%%%%%%%%%%%%%%%%%%%%%%%%%%%%%%% 

Single phase slips are observed in experimental situations which may be loosely
characterised as ``clean'', that is, for uncontaminated apertures of relatively
small sizes (a few micrometres at the most), with low background of mechanical and
acoustical interferences, {\it etc} ..., and with probing techniques that do not
manhandle the superfluid, namely, with low frequency Helmholtz resonators.  When
these conditions are not met, flow dissipation occurs in a more or less erratic
manner in large bursts -- multiple phase slips or `collapses' of the superflow.

Multiple phase slips and collapses constitute an apparent disruption of the
vortex nucleation mechanism described in the previous Section. Their
properties have been studied in detail in Ref.\cite{Avenel:95} and are briefly
mentioned below, together with possible mechanisms for their formation. It is
likely that these events provide a bridge between the ``clean'' single phase
slip case and the usual situation of the Feynman type critical velocities that
are temperature-independent below 1 K and dependent on the channel size. This
problem, which is not fully resolved at present, almost certainly involves some
form of preexisting vorticity.

\subsection{Remnant vorticity and vortex mills}

Remnant vorticity in $^4$He, which has long been assumed, has been shown
directly to exist by Awschalom and Schwarz by looking at the trapping of ions
by vortex lines \cite{Awschalom:84}. Vortices,
presumably nucleated at the \greeksym{l} transition where the critical
velocity is very low, remain stuck in various places of the superfluid sample
container. This trapped vorticity, according to Adams {\it et al.} \cite{Adams:85},
either is quite loosely bound to the substrate and disappears rapidly, or is
strongly pinned and is dislodged only by strong perturbations.

To account for laboratory observations and with the outcomes of numerous
numerical simulations of vortex dynamics, Schwarz has proposed the following
formula for the velocity at which vortices unpin \cite{Schwarz:85},
\begin{equation}        \label{Unpin}
v_\m u \lesssim  \frac{\kappa_4}{2\pi D}\,\ln\left(\frac{b}{a_0}\right) \;,
\end{equation}
$D$ being the size of the pinned vortex and $b$ being a characteristic size of
the pinning asperity. Equation (\ref{Unpin}) bears a strong resemblance with
that for the Feynman critical velocity, Eq.(\ref{FeynmanVelocity}). Long
vortices unpin at very low velocities unless they are perched on a tall
pedestal, but very small vortices pinned on microscopic defects at the cell
walls can in principle exist under a wide range of superflow velocities; a 200
nm long vortex filament pinned at both ends on 20 \AA\ asperities resists
transverse flows of velocities up to 10 cm/s.

In connection with the critical velocity problem, the long standing suggestion
by Glaberson and Donnelly \cite{Glaberson:66} of vortex mills still prevails.
In these authors'\,views, imposing a flow on a vortex pinned between the
opposite lips of an aperture would induce deformations such that the vortex
would twist on itself, undergo self-reconnections, and mill out free vortex
loops. First, we note that, according to Eq.(\ref{Unpin}), such a mill must
involve a pinned vortex of sub-micrometric size for any flow velocity above
$\sim\,$1 cm/s in order for the pinned vortex not to be washed away.  Thus, it
cannot as such account for Feynman-type critical velocities found in large
channels. Also, as shown by numerical simulations of 3D flows involving few
vortices only \cite{vortexstability}, vortex loops and filaments are stable
even against large deformations.  Vortices are not prone to twist on
themselves and foster loops.  It takes the complex flow fields associated with
fully developed vortex tangles to produce small rings
\cite{Svistunov:95,Tsubota:00}.  And it takes some quite special vortex
pinning geometry to set up a mill that actually works.

Schwarz has demonstrated the existence of such a mill by numerical simulations
\cite{Schwarz:90}. Let us take a vortex pinned at one end and floating along
the flow streamlines with its other end moving freely on the wall; this vortex
develops a helical motion, a sort of driven Kelvin wave, and reconnects
sporadically to the wall when the amplitude of the helical motion grows large
enough. This helical mill, which has to be of sub-micrometric size to stand
the flow, does churn out fresh vortices.

Vortex mills are thus unlikely to explain the critical velocities of the
Feynman type in the simple scheme suggested by Glaberson and Donnelly
\cite{Glaberson:66}.  However, the occurrence of multiple slips, which can be
seen in Fig.\,\ref{TimeChart}, is probably caused by some form of vortex mills
on a microscopic size.  Before coming to this topic, I need to describe
multiple slips in greater details.  But, at this point, the above remarks on
the stability of vortex loops or half-loops in their course already make it
unlikely that multiple slips be due to the production of small rings by the
nucleating vortices twisting on themselves à la Glaberson-Donnelly, as
suggested by Amar {\it et al.} \cite{Amar:92}. We have to dig a little further
to devise a scheme that works.

\subsection{The two types of large slips}

% Examples of multiple slips can be seen in Fig.\,\ref{TimeChart} which shows the
% peak amplitude chart of a two-aperture resonator at 12.5 mK, 24 bars, in a 100 ppb
% $^3$He in $^4$He sample. The very large amplitude drop shown in
% Fig.\,\ref{TimeChart} and in the insert is rare (one in $10^4 \mbox{ to } 10^5$
% slips) under the conditions of this particular experiment.  This type of events,
% called in \cite{Avenel:95} `singular' collapses and discussed further below, may occur at
% velocities much below the vortex nucleation threshold (down to at least a third of
% $v_c$).

Besides the usual single slip pattern, there appears in Fig.\,\ref{TimeChart}
occasional double slips (i.e.  involving phase changes by $4\pi$) and
infrequent triple slips.  Raising the temperature to 80 mK, again for this
particular cool-down, causes these multiple slips to occur much more frequently
and to involve more circulation quanta on the mean. These features are
described in detail in Ref.\cite{Avenel:95}.  As the probability for a
one-slip event per half-cycle is not large, that for a double slip is small,
and it becomes negligible for higher multiples. A separate mechanism for their
formation must be found.

% ---------------------------------------------------------------------------
\begin{figure}[t] 
  \begin{center}
    \includegraphics[angle=90,width=12 cm]{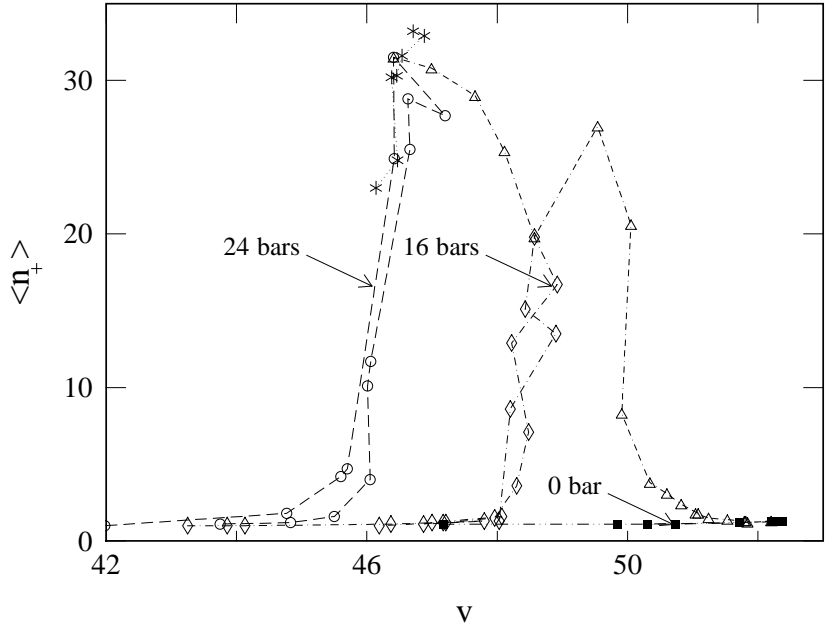}  % from qfs95-2.mac
    \caption{\label{MultiSlipFig} Mean size of (positive) multiple slips {\it
        vs} velocity in phase winding number in nominal purity $^4$He (100 ppb
      $^3$He): ($\scriptstyle\triangle$) pressure sweep from 0.4 to 24 bars at
      81.5 mK (all even values of the pressure $P$, and 0.4, 1, 3, 5, 7 bars)
      -- ($\diamond$) temperature sweep at 16 bars -- ($\circ$) temperature
      sweep at 24 bars - ($\ast$) drive level sweep at 24 bars, 81.5 mK --
      ($\scriptscriptstyle\blacksquare$) temperature sweep at 0 bar. For the
      temperature sweeps, from 14 to 200 mK approximately, $v$ first
      increases, reaches the quantum plateau and then decreases, as shown in
      the insert of Fig.\,\protect{\ref{vc}}.  Lines connect successive data
      points in the temperature and pressure sweeps. }
  \end{center}              
\end{figure}
% ---------------------------------------------------------------------------

Some degree of understanding of the formation of multiple slips can be gained
by plotting the mean value of the phase slip sizes, expressed in number of
quanta, against the flow velocity at which the slips take place
\cite{Varoquaux:95}. This flow velocity is close to the critical velocity for
single phase slips, {\it i.e.} the vortex nucleation velocity; it varies with
temperature, pressure, and resonator drive level. A plot summarising these
variations is shown in Fig.\,\ref{MultiSlipFig} for $<\!n_{_{+}}\!\!>$, {\it
  i.e.}  in the flow direction conventionally chosen as the $+$ direction.
Slips in the opposite $(-)$ direction behave qualitatively in the same manner
but the phenomenon displays a clear quantitative asymmetry. As can be seen in
Fig.\,\ref{MultiSlipFig}, the mean slip size decreases, as does the nucleation
velocity, on either side of the quantum plateau -- a $^3$He impurity effect on
the low $T$ side -- a thermal effect on the high-$T$ side.  However, it
increases with pressure, contrarily to the nucleation velocity which decreases
with increasing pressure.

We conclude from the organisation of the data with the various parameters in
Fig.\,\ref{MultiSlipFig} that the magnitude of the superflow velocity does not
directly control, by itself, the occurrence of multiple slips. In turn, this
implies, as will be discussed further below, that the phenomenon under study
is not purely ruled by hydrodynamics in the bulk of the fluid but involves
some complex interplay with the boundaries.  As shown in
Fig.\,\ref{MultiSlipFig}, the velocity threshold for the appearance of
multiple slips depends on hydrostatic pressure; in fact, the $P$-dependence of
the upturn of $<\!n_{_{+}}\!\!>$ vs $v$ exactly tracks that of the critical
velocity for single phase slip nucleation. This indicates that multiple slips
appear because of an alteration, or as a consequence, of the nucleation
process itself.

The pattern of formation of multiple slips changes from cool-down of the cell from
room temperature to cool-down but remains stable for each given cool-down. It seems
to depend on the degree of contamination of the cell, degree which cannot easily
be controlled experimentally. The detailed microscopic configuration of the
aperture wall where nucleation takes place probably plays an major role in
multiple slip formation. 

Another kind of very large drops in the resonance amplitude of the resonator
was also observed, which sometimes resulted in a complete collapse of the
resonance. These ``singular'' collapses, first observed by Hess
\cite{Hess:77}, occur at flow velocities that are lower than the critical
velocity for phase slips, sometimes as low as $v_\m c/3$. Multiple slips are
different from ``singular'' collapses and the underlying mechanisms
responsible for both phenomena are bound to be different, as discussed below.
  
\subsection{{\it In-situ} contamination by atomic clusters: pinning
  and collapses} 

In a series of experiments conducted at Saclay
\cite{Hakonen:98b,Varoquaux:98}, in which the experimental cell was
deliberately heavily contaminated by atomic clusters of air or H$_2$, we
observed that numerous multiple slips and collapses of the ``singular'' type
occurred. The peak amplitude charts of the resonator became mostly impossible
to interpret, except in a few instances where two apparent critical velocities
for single phase slips were observed. The higher critical velocity corresponds
to the one observed in the absence of contamination.  The lower critical
velocity is thought to reveal the influence of a vortex pinned in the
immediate vicinity of the nucleation site.  This vortex induces a local
velocity which adds to that of the applied flow and causes an apparent
decrease in the critical velocity for phase slips. Because of this change, the
presence of the pinned vortex could be monitored, the lifetime in the pinned
state and the unpinning velocity could be measured, yielding precious
information on the pinning process.

This observation, reported in detail in Refs.\cite{Hakonen:98b} and
\cite{Varoquaux:98}, shows that pinned vorticity can contribute to the
nucleation of new vortices at the walls of the experimental cell. Such pinned
vortices as the one described above can, instead of interacting with the
nucleation site, set up a transient vortex mill of the helical type and
generate a burst of vortices. The existence of such pinned vortices is
established; that they can form a micro-mill is highly plausible. We thus have
a possible explanation for multiple slip formation \cite{Varoquaux:01}.  The
pinning event would take place immediately after nucleation when the velocity
of the vortex relative to the boundary is still very small and the capture by
a pinning site easy. The micro-mill remains in activity as long as the flow is
sufficient to maintain the helical instability, which depends on the pinning
stand geometry. As it is set up to withstand one flow direction, it is
destroyed when the flow velocity reverses itself in the resonance motion. It
eventually re-establishes itself during a subsequent resonance cycle, causing
a new multiple slip.  This process depends on the precise details of the
pinning site configuration and of the primordial vortex trajectory, factors
which allow for the variableness of multiple slips on contamination and
pressure.

In the same experiments, we observed that a large number of unpinning events
were taking place at an ``anomalously low'' unpinning velocity. A parallel can
be made \cite{Varoquaux:98} with the singular collapses that also occur at
``subcritical'' velocities. In fact, both phenomena were seen quite frequently
in these experiments, suggesting that they have a common cause.
Noting furthermore that pinning and unpinning processes were quite
frequent, releasing a fair amount of vagrant vorticity, it appears quite
plausible that both singular collapses and low velocity unpinning events are
caused by vagrant vortices hopping from pinning sites to pinning sites,
ending up passing by close to a pinned vortex or a vortex nucleation site,
and giving a transient boost to the local velocity, which pushes a pinned
vortex off its perch or causes a burst of vortices to be shed.

These observations, albeit incidental, have important consequences for the
critical velocity problem: existing vortices, either pinned or free-moving,
can contribute to the nucleation of new vortices at the walls of the
experimental cell at apparent velocities much lower than the critical velocity
for phase slips.  We are thus provided with a mechanism by which superflow
dissipation sets in at mean velocities on the large scale much smaller the
velocity for vortex nucleation on the microscopic scale, possibly bridging the
gap between phase slip and Feynman type critical velocities.  Vortex
nucleation at the walls is also quite likely to take part in the build-up of
self-sustaining vortex tangles forming superfluid turbulence, up to now
attributed solely to reconnection mechanisms \cite{Schwarz:83}.

To conclude, the critical velocities in superfluids that are true and proven
include the Landau critical velocity for roton creation, the formation of
vortices by a hydrodynamical instability in BEC gases \cite{Madison:01} and in
$^3$He \cite{Eltsov:05}, the nucleation of vortices by thermal activation and
quantum tunnelling in $^4$He, both for ion propagation and in aperture flow. I
have presented rather compelling experimental evidence for the interplay
between vortex nucleation and pinned vorticity on a microscopic scale; this
evidence points toward the existence of helical vortex micro-mills that can
generate vortices at fairly low applied velocities. Finally, vagrant vortices
interacting with these mills, or with vortex nucleation sites, are found to
generate enough vorticity to completely kill the applied superflow and explain
singular collapses. The study of phase slippage has taken us a long way toward
an explanation of critical velocities in superfluid helium-4.

% ***************************************************************************

\newcommand{\noopsort}[1]{}


\begin{thebibliography}{10}
\expandafter\ifx\csname url\endcsname\relax
  \def\url#1{\texttt{#1}}\fi
\expandafter\ifx\csname urlprefix\endcsname\relax\def\urlprefix{URL }\fi

\bibitem{Wilks:67}
J.~Wilks, The properties of Liquid and Solid Helium, Clarendon Press - Oxford,
  1967.

\bibitem{Vinen:63}
W.~Vinen, Critical velocities in liquid helium {II}, in: Liquid Helium, ed. G.
  Careri, Academic Press - New-York, 1963, p. 336.

\bibitem{Langer:70}
J.~Langer, J.~Reppy, Prog. Low Temp. Phys., Vol.~6, ed. C. J. Gorter,
  North-Holland - Amsterdam, 1970, Ch.~1, p.~1.

\bibitem{Muirhead:84}
C.~Muirhead, W.~Vinen, R.~Donnelly, The nucleation of vorticity by ions in
  superfluid $^4${H}e i. basic theory, Phil. Trans. Roy. Soc. A 311 (1984) 433.

\bibitem{Varoquaux:87}
E.~Varoquaux, O.~Avenel, M.~Meisel, Phase slippage and vortex nucleation in the
  critical flow of superfluid $^4${H}e through an orifice, Can. J. Phys. 65
  (1987) 1377.

\bibitem{Varoquaux:91}
E.~Varoquaux, W.~{Zimmermann, Jr.}, O.~Avenel, Phase-slippage studies of the
  critical velocity in helium-4, in: Excitations in Two-Dimensional and
  Three-Dimensional Quantum Fluids, eds. A. F.G. Wyatt and H. J. Lauter, Plenum
  press, New-York, 1991, p. 343.

\bibitem{Avenel:93}
O.~Avenel, G.~Ihas, E.~Varoquaux, The nucleation of vortices in superfluid
  $^4${H}e: Answers and questions, J. Low Temp. Phys. 93 (1993) 1031.

\bibitem{Varoquaux:01}
E.~Varoquaux, O.~Avenel, Y.~Mukharsky, P.~Hakonen, The experimental evidence
  for vortex nucleation in $^4${H}e, in: W.~V. C.F.~Barenghi, R.J.~Donnelly
  (Ed.), Quantized Vortex Dynamics and Superfluid Turbulence, Springer -
  Berlin, 2001, p.~36.

\bibitem{Varoquaux:03}
E.~Varoquaux, O.~Avenel, Vortex nucleation in phase-slippage experiments in
  ultrapure superfluid $^4${H}e below 0.5 {K}, Phys. Rev. B68 (2003) 054515.

\bibitem{Baym:69}
G.~Baym, Mathematical Methods in Solid State and Superfluid Theory, eds. R.C.
  Clark and G.H. Derrick, Oliver and Boyd Ltd, Edinburgh, 1969, Ch.~3, p. 134.

\bibitem{Anderson:66a}
P.~Anderson, Considerations on the flow of superfluid helium, Rev. Mod. Phys.
  38 (1966) 298.

\bibitem{Lamb:32}
S.~H. Lamb, Hydrodynamics, sixth edition Edition, Cambridge University Press -
  Cambridge, 1932.

\bibitem{Sonin:87}
E.~Sonin, Vortex oscillations and hydrodynamics of rotating superfluids, Rev.
  Mod. Phys. 59 (1987) 87.

\bibitem{Sonin:95}
E.~Sonin, Nucleation and creep of vortices in superfluids and clean
  superconductors, Physica B 210 (1995) 234.

\bibitem{Avenel:85}
O.~Avenel, E.~Varoquaux, Observation of singly quantized dissipation events
  obeying the {J}osephson frequency relation in the critical flow of superfluid
  $^4${H}e through an aperture, Phys. Rev. Lett. 55 (1985) 2704.

\bibitem{Shifflett:92}
G.~Shifflett, G.~Hess, Intrinsic critical velocities in superfluid $^4${H}e
  flow through 12-$\mu$m diam. orifices near {$T_{\lambda}$}: Experiments on
  the effect of geometry, J. Low Temp. Phys. 98 (1995) 591.

\bibitem{Zimmermann:96}
W.~{Zimmermann, Jr.}, {The flow of superfluid $^4$He through submicron
  apertures: phase slip and critical velocities due to quantum vortex motion},
  Contemp. Phys. 37 (1996) 219.

\bibitem{Packard:98}
R.~Packard, The role of the {J}osephson-{A}nderson equation in superfluid
  helium, Rev. Mod. Phys. 70 (1998) 641.

\bibitem{Avenel:87}
O.~Avenel, E.~Varoquaux, Josephson effect and phase slippage in superfluids,
  Jpn. J. Appl. Phys. 26 (1987) 1798.

\bibitem{Beecken:87}
B.~P. Beecken, W.~{Zimmermann, Jr}., Variation of the critical order-parameter
  phase difference with temperature from 0.4 to 1.9 {K} in the flow of
  superfluid $^4${H}e through a tiny orifice, Phys. Rev. 35 (1987) 1630.

\bibitem{Varoquaux:94}
E.~Varoquaux, O.~Avenel, Phase slip phenomena in superfluid helium, Physica B
  197 (1994) 306.

\bibitem{Avenel:95}
O.~Avenel, M.~Bernard, S.~Burkhart, E.~Varoquaux, Autopsy of superflow
  collapses and multiple phase slips, Physica B 210 (1995) 215.

\bibitem{SQUID}
SQUID is an acronym for Superconducting QUantum Interferometric Device. The
  present sensitivity of the displacement sensor is $\sim 10^{-15}$ m.

\bibitem{winding}
Phase winding numbers are related to velocities in cm/s by multiplication by
  $l_{\text h}/\kappa_4$, the ``hydraulic'' length $l_{\text h}$ characterising
  the geometry of the aperture. For a phase slip by 2$\pi$, the phase winding
  number changes by one unit and the trapped circulation in the resonator loop
  by one quantum.

\bibitem{Burkhart:94}
S.~Burkhart, M.~Bernard, O.~Avenel, E.~Varoquaux, Scenario for a quantum phase
  slip, Phys. Rev. Lett. 72 (1994) 380.

\bibitem{Josserand:95}
C.~Josserand, Y.~Pomeau, {Generation of vortices by the Kadomtsev-Petviashvili
  instability}, Europhys. Lett. 30 (1995) 43.

\bibitem{Josserand:95b}
C.~Josserand, Y.~Pomeau, S.~Rica, Cavitation versus vortex nucleation in a
  superfluid model, Phys. Rev. Lett. 75 (1995) 3150.

\bibitem{Andreev:04}
A.~Andreev, L.~Melnikovsky, Thermodynamics of superfluidity, J. Low Temp. Phys.
  135 (2004) 411.

\bibitem{Hulin:74}
J.~Hulin, D.~D'Humi\`eres, B.~Perrin, A.~Libchaber, Critical velocities for
  superfluid helium flow through a small hole, Phys. Rev. A9 (1974) 885.

\bibitem{Zimmermann:93}
W.~{Zimmermann, Jr.}, The critical velocity behavior of superfluid $^4${H}e in
  a micron-size aperture at two different frequencies of oscillatory flow, J.
  Low Temp. Phys. 91 (1993) 219.

\bibitem{McClintock:91}
P.~McClintock, R.~Bowley, Vortex creation in superfluid helium-4, in:
  Excitations in Two-Dimensional and Three-Dimensional Quantum Fluids, eds. A.
  F.G. Wyatt and H. J. Lauter, Plenum press, New-York, 1991, p. 567.

\bibitem{McClintock:95}
P.~McClintock, R.~Bowley, The {L}andau critical velocity, in: W.~Halperin
  (Ed.), Prog. Low Temp. Physics, Vol. XIV, Elsevier, 1995, p.~1.

\bibitem{Rayfield:64}
G.~W. Rayfield, F.~Reif, Quantized vortex rings in superfluid helium, Phys.
  Rev. 136 (1964) A1194.

\bibitem{attempt}
A noteworthy attempt to by-pass this experimental finding is that of
  Ref.\protect\cite{Andreev:04}.

\bibitem{Zimmermann:98}
W.~{Zimmermann, Jr.}, C.~Lindensmith, J.~Flaten, The interpretation of
  phase-slip and critical-velocity data ..., J. Low Temp. Phys. 110 (1998) 497.

\bibitem{Varoquaux:93}
E.~Varoquaux, G.~Ihas, O.~Avenel, R.~Aarts, Vortex nucleation in superfluid
  $^4${H}e probed by $^3${H}e impurities, Phys. Rev. Lett. 70 (1993) 2114.

\bibitem{Steinhauer:95}
J.~Steinhauer, K.~Schwab, Y.~Mukharsky, J.~Davis, R.~E. Packard, Vortex
  nucleation in superfluid $^4${H}e, Phys. Rev. Lett. 74 (1995) 5056.

\bibitem{Zimmermann:90}
W.~{Zimmermann, Jr.}, O.~Avenel, E.~Varoquaux, Critical flow of superfluid
  helium-4 through a submicron aperture: width of the critical transition,
  Physica B 165\&166 (1990) 749.

\bibitem{Varoquaux:86}
E.~Varoquaux, M.~Meisel, O.~Avenel, Onset of the critical velocity regime in
  superfluid $^{4}${H}e at low temperature, Phys. Rev. Lett. 57 (1986) 2291.

\bibitem{Nature}
``Nature is trying to tell us something'', an idiom often used by Douglas
  Osheroff in connection with helium physics.

\bibitem{Ihas:92}
G.~Ihas, O.Avenel, R.~Aarts, R.~Salmelin, E.~Varoquaux, Quantum nucleation of
  vortices in the flow of superfluid $^4${H}e through an orifice, Phys. Rev.
  Lett. 69 (1992) 327.

\bibitem{Hendry:88}
P.~Hendry, N.~Lawson, P.~McClintock, C.~Williams, R.~Bowley, Macroscopic
  quantum tunneling of vortices in {H}e {II}, Phys. Rev. Lett. 60 (1988) 604.

\bibitem{Martinis:88}
J.~Martinis, H.~Grabert, Thermal enhancement of macroscopic quantum tunneling:
  Derivation from noise theory, Phys. Rev. B 38 (1988) 2371.

\bibitem{Caldeira:83a}
A.~Caldeira, A.~Leggett, Quantum tunnelling in a dissipative system, Ann. Phys.
  (N.Y.) 149 (1983) 374, erratum, {153} (1984) 445(E).

\bibitem{Landau:barrier}
L.~Landau, E.~Lifshitz, Quantum Mechanics, Pergamon Press - London, 1958,
  1\raisebox{0.8ex}{\tiny{rst}} edition \S 50.

\bibitem{Coleman:77}
S.~Coleman, Fate of the false vacuum: Semiclassical theory, Phys. Rev. D 15
  (1977) 2929.

\bibitem{Fischer:00}
U.~Fischer, Tunnelling of topological line defects in strongly coupled
  superfluids, Ann. Phys. (Leipzig) 9 (2000) 523.

\bibitem{Larkin:84}
A.~Larkin, K.~Likharev, Y.~Ovchinnikov, Secondary quantum macroscopic effects
  in weak superconductivity, Physica B 126 (1984) 414.

\bibitem{Frisch:92}
T.~Frisch, Y.~Pomeau, S.~Rica, Transition to dissipation in a model of
  superflow, Phys. Rev. Lett. 69 (1992) 1644.

\bibitem{Nore:00}
C.~Nore, C.~Huepe, M.~Brachet, Subcritical dissipation in three-dimensional
  superflows, Phys. Rev. Lett. 84 (2000) 2191.

\bibitem{Berloff:01}
{N.G. Berloff and P.H. Roberts}, Vortices in nonlocal condensate models of
  superfluid helium, in: Quantized Vortex Dynamics and Superfluid Turbulence,
  2001, p. 268, {\it loc.cit.}

\bibitem{Rica:01}
S.~Rica, Vortex nucleation and limit speed for a flow passing nonlinearly
  around a disk in the nonlinear {S}chrödinger equation, Vol. 571 of Lecture
  Notes in {Physics}, Springer, Berlin, 2001, p. 258.

\bibitem{Melnikov:91}
V.~Mel'nikov, The {K}ramers problem: Fifty years of development, Phys. Reports
  209 (1991) 1.

\bibitem{further}
See Ref.\protect\cite{Varoquaux:03} for further justifications and references
  to the literature.

\bibitem{Martinis:87}
J.~M. Martinis, M.~H. Devoret, J.~Clarke, {Experimental tests for the quantum
  behavior of a macroscopic degree of freedom: The phase difference across a
  Josephson junction}, Phys. Rev. B 35 (1987) 4682.

\bibitem{Grabert:87}
H.~Grabert, P.~Olschowski, U.~Weiss, Quantum decay rates for dissipative
  systems at finite temperarures, Phys. Rev. B36 (1987) 1931.

\bibitem{Waxman:85}
D.~Waxman, A.~Leggett, Dissipative quantum tunneling at finite temperatures,
  Phys. Rev. B 32 (1985) 4450.

\bibitem{Grabert:88}
H.~Grabert, Escape from a metastable well: The {K}ramers turnover problem,
  Phys. Rev. Lett. 61 (1988) 1683.

\bibitem{Rips:89}
{I. Rips and E. Pollak}, Quantum {K}ramers model: Solution of the turnover
  problem, Phys. Rev. A 41 (1989) 5366.

\bibitem{Hanggi:90}
P.~H\"anggi, P.~Talkner, M.~Borkovec, Reaction rate theory: Fifty years after
  {K}ramers, Rev. Mod. Phys. 62 (1990) 251.

\bibitem{Langer:67b}
J.~Langer, M.~Fisher, Intrinsic critical velocity of a superfluid, Phys. Rev.
  Lett. 19 (1967) 560.

\bibitem{Volovik:72}
G.~Volovik, Quantum-mechanical formation of vortices in a superfluid liquid,
  Sov. Phys. JETP Lett. 15 (1972) 81.

\bibitem{Awschalom:84}
D.~Awschalom, K.~W. Schwarz, Observation of a remanent vortex-line density in
  superfluid helium, Phys. Rev. Lett. 52 (1984) 49.

\bibitem{Adams:85}
P.~Adams, M.~Cieplak, W.~I. Glaberson, Spin-up problem in superfluid $^4${H}e,
  Phys. Rev. B 32 (1985) 171.

\bibitem{Schwarz:85}
K.~Schwarz, Three-dimensional vortex dynamics in superfluid $^4${H}e: Line-line
  and line-boundary interactions, Phys. Rev. B 31 (1985) 5782.

\bibitem{Glaberson:66}
W.~Glaberson, R.~Donnelly, Growth of pinned quantized vortex lines in helium
  {II}, Phys. Rev. 141 (1966) 208.

\bibitem{vortexstability}
K.W. Schwarz: private communication.

\bibitem{Svistunov:95}
B.~Svistunov, Superfluid turbulence in the low-temperature limit, Phys. Rev. B
  52 (1995) 3647.

\bibitem{Tsubota:00}
M.~Tsubota, T.~Araki, S.~K. Nemirovskii, Dynamics of vortex tangle without
  mutual friction in superfluid $^4${H}e, Phys. Rev. B 62 (2000) 11751.

\bibitem{Schwarz:90}
K.~Schwarz, Phase slip and turbulence in superfluid $^4${H}e: A vortex mill
  that works, Phys. Rev. Lett. 64 (1990) 1130.

\bibitem{Amar:92}
A.~Amar, Y.~Sasaki, R.~Lozes, J.~Davis, R.~Packard, Quantized phase slippage in
  superfluid $^4${H}e, Phys.Rev. Lett. 68 (1992) 2624.

\bibitem{Varoquaux:95}
E.~Varoquaux, O.~Avenel, M.~Bernard, S.~Burkhart, Multiple quantum phase slips
  in superfluid $^4${H}e, J. Low Temp. Phys. 101 (1995) 821.

\bibitem{Hess:77}
G.~Hess, Vortex generation in modulated superfluid $^4${H}e flow through a
  pinhole, Phys. Rev. B 15 (1977) 5204.

\bibitem{Hakonen:98b}
P.~Hakonen, O.~Avenel, E.~Varoquaux, Evidence for single-vortex pinning and
  unpinning events in superfluid $^4${H}e, Phys. Rev. Lett. 81 (1998) 3451.

\bibitem{Varoquaux:98}
E.~Varoquaux, O.~Avenel, P.~Hakonen, Y.~Mukharsky, Observation of single-vortex
  pinning in superfluid $^4${H}e, Physica B 255 (1998) 55.

\bibitem{Schwarz:83}
K.~Schwarz, Critical velocity for a self-sustaining vortex tangle in superfluid
  helium, Phys. Rev. Lett. 50 (1983) 364.

\bibitem{Madison:01}
K.~W. Madison, F.~Chevy, V.~Bretin, J.~Dalibard, {Stationary States of a
  Rotating Bose-Einstein Condensate: Routes to Vortex Nucleation}, Phys. Rev.
  Lett. 86 (2001) 4443.

\bibitem{Eltsov:05}
V.~Eltsov, M.~Krusius, G.~Volovik, {Vortex Formation and Dynamics in Superfluid
  $^3$He}, in: W.~Halperin (Ed.), Prog. Low Temp. Physics, Vol.~XV, Elsevier,
  2005, Ch.~1, p.~1.

\end{thebibliography}
\end{document}